\documentclass[twocolumn,english,twocolum,superscriptaddress,natbib,amsmath,amssymb,floatfix,nofootinbib]{revtex4-2}
\usepackage[colorlinks=true,citecolor=blue,linkcolor=magenta]{hyperref}
\usepackage{xcolor}
\usepackage{amsmath}
\usepackage{amsfonts}
\usepackage{stmaryrd}
\usepackage{amssymb}
\usepackage{graphicx}
\usepackage{dcolumn}
\usepackage{multirow}
\usepackage{bm}
\usepackage{siunitx}
\usepackage{xurl}
\usepackage[version=4]{mhchem}
\usepackage{float}
\usepackage[colorlinks=true,citecolor=blue,linkcolor=magenta]{hyperref}
\usepackage[english]{babel}
\usepackage{url}
\DeclareSIUnit \dbc {dBc}

\newcommand{\ErSiN}{Er:Si$_3$N$_4$}

\newcommand{\TmAlO}{Tm:Al$_2$O$_3$}
\newcommand{\SiN}{Si$_3$N$_4$}

\newcommand{\epflaff}{\affiliation{\mbox{Institute of Physics}, \mbox{Swiss Federal Institute of Technology Lausanne (EPFL)}, CH-1015 Lausanne, Switzerland}
	\affiliation{\mbox{Institute of Electrical and Microengineering}, \mbox{Swiss Federal Institute of Technology Lausanne (EPFL)},\\ CH-1015 Lausanne, Switzerland}}
\newcommand{\hzdraff}{\affiliation{Helmholtz-Zentrum Dresden-Rossendorf (HZDR), 01328 Dresden, Germany}}
\newcommand{\cofirst}{\textsuperscript{\textdagger}}
\newcommand\blfootnote[1]{
	\begingroup
	\renewcommand\thefootnote{}\footnote{#1}%
	\addtocounter{footnote}{-1}%
	\endgroup
}

\begin{document}

\setlength{\parskip}{6pt}%
	
\title{High-pulse-energy integrated mode-locked lasers based on a Mamyshev oscillator}

\author{Zheru Qiu\cofirst}\epflaff
\author{Jianqi Hu\cofirst}\epflaff
\author{Xuan Yang\cofirst}\epflaff
\author{Zhongshu Liu}\epflaff
\author{Yichi Zhang}\epflaff
\author{Xinru Ji}\epflaff
\author{Jiale Sun}\epflaff
\author{Grigorii Likhachev}\epflaff
\author{Xurong Li}\epflaff
\author{Zihan Li}\epflaff
\author{Ulrich Kentsch}\hzdraff
\author{Tobias J. Kippenberg}
\email{tobias.kippenberg@epfl.ch}
\epflaff
\medskip
\maketitle
\blfootnote{\hspace{-0.8em}\cofirst~These authors contributed equally.}
\noindent\textbf{%
Ultrafast lasers have unlocked numerous advances across science and technology: they enable corneal surgery~\cite{juhasz2002corneal}, reveal chemical reaction dynamics~\cite{zewail2000femtochemistry}, and underpin optical atomic clocks~\cite{diddams2001optical}.
Over the past decades, extensive efforts have been devoted to developing photonic integrated circuit-based mode-locked lasers that are compact, scalable, and compatible with further on-chip functionalities~\cite{byun2009integrated,lownoisehetero2021,guo2023ultrafast}.
Yet, existing implementations fall short of pulse energies required for their subsequent uses in nonlinear applications.
In this work, we demonstrate the first mode-locked laser that overcomes this limitation in low-loss erbium-doped silicon nitride photonic integrated circuits~\cite{liu2022photonic}.
The laser is based on the Mamyshev oscillator architecture~\cite{regelskis2015ytterbium,liu2017megawatt}, which employs alternating spectral filtering and self-phase modulation for mode-locking.
It delivers a 176~MHz stream of pulses with nanojoule energy, comparable to fiber lasers and surpassing previous photonic integrated sources by more than two orders of magnitude.
The output pulses exhibit excellent coherence, can be linearly compressed to 147~fs and directly drive a 1.5-octave-spanning supercontinuum in an integrated waveguide.
Our work establishes a new generation of high-pulse-energy photonic integrated mode-locked lasers and paves the way for their widespread adoption. 
}

\section{Introduction}
\begin{figure*}
	\centering
	\includegraphics[width=1\linewidth]{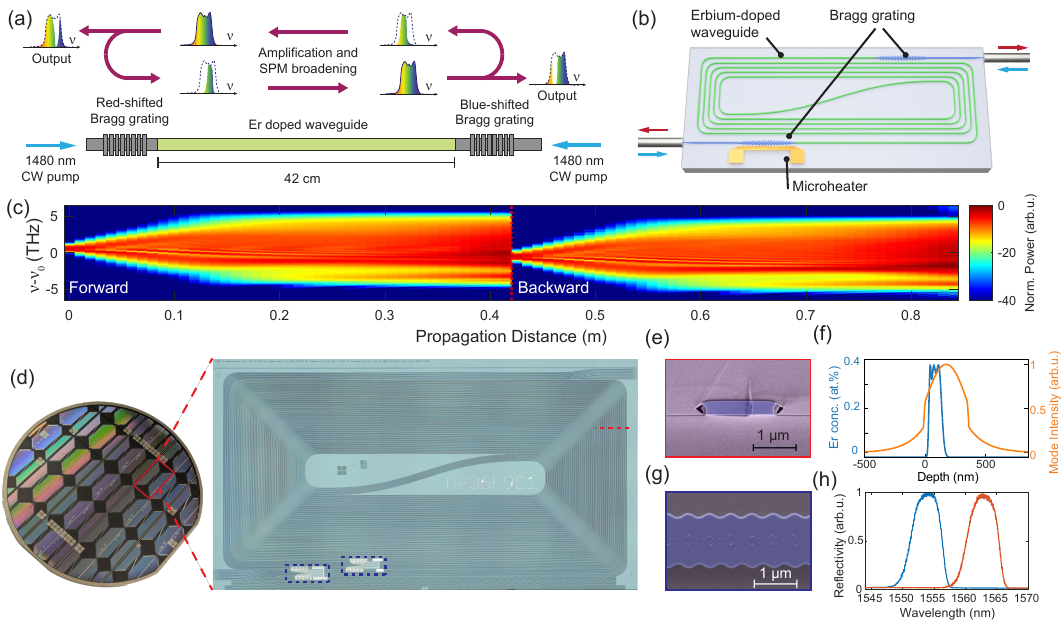}
	\caption{\textbf{Principle and fabrication of the integrated mode-locked lasers based on Mamyshev oscillator.}
		(a) Illustration of the working principle of the integrated Mamyshev oscillator, consisting of a 42~cm \ErSiN{} waveguide sandwiched between two spectrally offset WBGs. 
		(b) Schematic of the integrated Mamyshev oscillator, showing the erbium-doped waveguide arranged in a spiral as well as the WBGs with integrated microheater for tuning.
		(c) Numerically simulated optical spectrum of the Mamyshev oscillator as a function of propagation path length $l$ in the linear cavity.  
		(d) Photograph of the fabricated wafer and a chip that contains 26 individual MLLs.
		The blue boxes indicate the location of the WBGs.
		The internal tracking number for the MLL sample used in the experiment is \texttt{D20602F8C2}.
		(e) False-colored scanning electron microscope (SEM) image of the doped waveguide cross-section.
		(f) Simulated fundamental transverse electric (TE) optical mode profile and the erbium concentration in the \SiN{} waveguide.
		(g) False-colored SEM image of the corrugated WBG.
		(h) Overlaid reflection spectra of the two WBGs, showing the spectral offset between the passbands.
	}
	\label{fig1}
\end{figure*}

Mode-locked lasers (MLLs) are optical sources that generate trains of intense, ultrashort pulses by locking the phases of different longitudinal modes~\cite{haus2002mode,keller2003recent}.
Since their inception, MLLs have become indispensable tools across numerous fields, enabling advances in material processing, metrology~\cite{udem2002optical}, biological imaging~\cite{xu2013recent}, ophthalmic surgery~\cite{juhasz2002corneal}, femtochemistry~\cite{zewail2000femtochemistry}, and optical spectroscopy.
Recently, significant efforts have been devoted to developing ultrafast pulse sources on photonic integrated circuits (PICs), aiming to replicate tabletop laser performance within a compact footprint and allowing for integration of additional functionalities.
For example, mode-locking of integrated III-V semiconductor lasers has been demonstrated both monolithically~\cite{lu2008312,moskalenko2014record} and through heterogeneous integration with silicon or lithium niobate photonics, leveraging saturable absorption in semiconductors~\cite{wang2017iii,liu2019high,Hermans21,lownoisehetero2021} or radio-frequency modulation~\cite{guo2023ultrafast,ling2024electrically}.
Using external semiconductor saturable absorber mirrors, mode-locking has also been achieved in low-confinement erbium-doped planar lightwave circuits~\cite{byun2009integrated}. 
Alternatively, Kerr microcombs~\cite{brasch2016photonic,helgason2023surpassing} and electro-optic frequency combs~\cite{yu2022integrated} were shown to generate femtosecond pulses on chip.
However, all existing PIC-based ultrafast sources still fall far short of the performance of tabletop MLL, in terms of noise, peak power, pulse duration, and pulse energy.
In particular, pulse energies in existing works remain at the few-picojoule level, limiting their ability to drive on-chip nonlinear processes such as supercontinuum generation, a cornerstone for self-referenced, phase-coherent links between optical and radio frequencies~\cite{udem2002optical}.

Recent advances in high-confinement doped waveguides offer a promising route to scaling up pulse energy.
Demonstrations of high-power optical amplifiers in erbium-ion-implanted silicon nitride (\ErSiN)~\cite{liu2022photonic}, thulium-doped aluminum oxide (\TmAlO)~\cite{singh2025watt}, and titanium-doped sapphire waveguides~\cite{wang2023photonic, yang2024titanium} highlight their potential as efficient gain media for integrated MLLs.
These low-loss doped waveguides also support sub-meter-long laser cavities within compact chips, enabling low repetition rates of a few hundred megahertz, which are crucial for achieving high pulse energy at a finite average power.
Yet, stable mode-locking has not been attained with these active waveguides, with pioneering prior demonstrations limited to Q-switched operation~\cite{shtyrkova2019integrated,singh2024silicon}.

To overcome the limitations of current integrated MLLs, we consider alternative mode-locking mechanisms~\cite{fu2018several}.
The Mamyshev oscillator stands out as a particularly promising approach for mode locking~\cite{regelskis2015ytterbium,liu2017megawatt}, allowing for record megawatt peak powers~\cite{Liu19} and few-cycle pulses~\cite{ma2019ultrabroadband} in fiber-based lasers.
The concept builds on the Mamyshev regenerators, originally proposed by P.V. Mamyshev for optical signal regeneration in soliton transmission, which exploits self-phase modulation and spectral filtering to produce a nonlinear transfer function~\cite{mamyshev1998all}.
By concatenating two such regenerators, a Mamyshev oscillator cavity can be formed with a nonlinear waveguide and two spectrally offset bandpass filters (Fig.\,\ref{fig1}(a)), as first demonstrated in~\cite{pitois2008generation,4591476}.
Unlike conventional mode-locking schemes, Mamyshev oscillators eliminate the need for a physical saturable absorber~\cite{regelskis2015ytterbium,liu2017megawatt}.
Mode-locking instead arises from a combination of nonlinear broadening and filtering: low-power light is suppressed by the non-overlapping filters, while high-power pulses broaden spectrally in the nonlinear segment and pass through both filters to sustain lasing.
A key advantage of the Mamyshev oscillator is its intrinsic tolerance to large intracavity nonlinear phase shifts as high as $>60\pi$~\cite{liu2017megawatt}, which can otherwise cause pulse breakup in conventional MLLs~\cite{grelu2012dissipative}.
This challenge is particularly pronounced in high-confinement integrated waveguides, where their effective nonlinearities are three orders of magnitude higher than fibers.
Furthermore, the Mamyshev oscillator only requires two critical components: nonlinear waveguides and bandpass filters, while the latter can be simply implemented on-chip using waveguide Bragg gratings (WBGs). 
This eliminates the need to integrate semiconductor saturable absorbers or engineer artificial ones, significantly reduces the complexity of integrated MLLs.

In this work, we demonstrate a photonic integrated Mamyshev oscillator based on erbium-doped silicon nitride waveguides, achieving record performance in chipscale MLLs (see Supplementary Information S1).
The laser generates coherent ultrafast nanojoule-level pulses, capable of directly driving octave-spanning supercontinuum in dispersion-engineered \SiN{} waveguides without amplification.

\section{Design and fabrication}

\begin{figure*}
	\centering
	\includegraphics[width=1\linewidth]{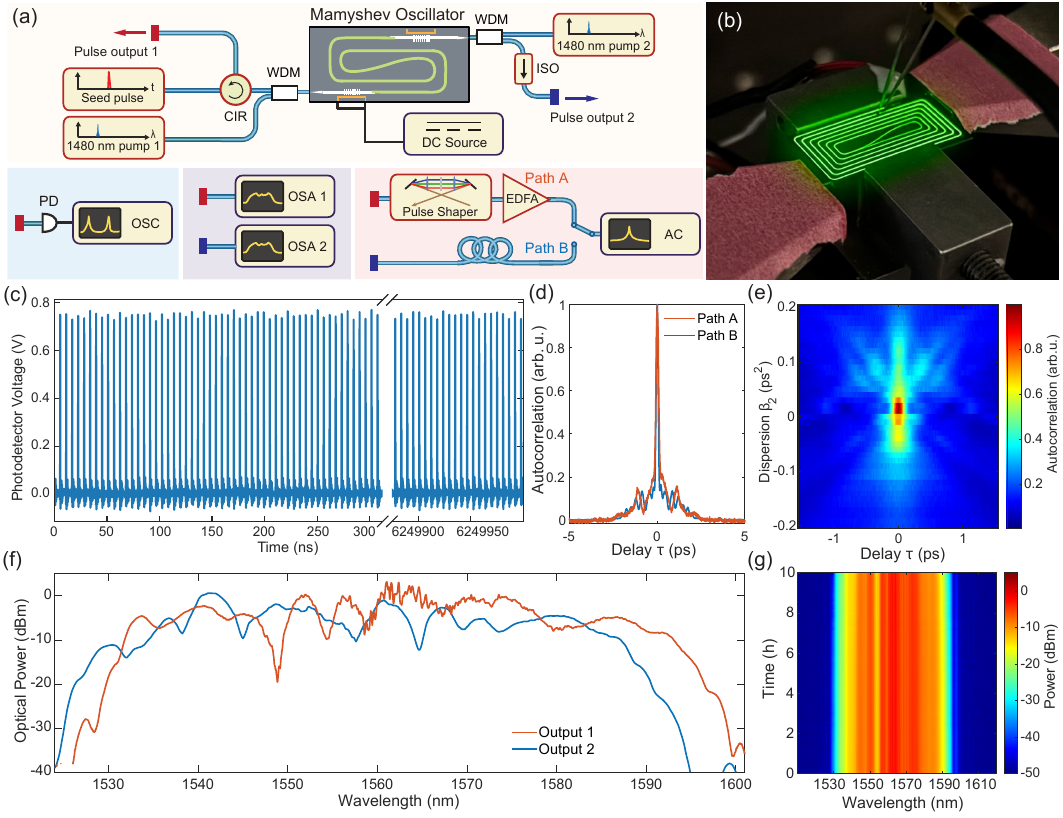}
	\caption{\textbf{Characterization of a photonic integrated Mamyshev oscillator.} 
		(a) Experimental setup for integrated MLL operation and characterization.
		The integrated Mamyshev oscillator is pumped by 1480~nm light through the WBGs and initiated by an external seed pulse.
		Optical spectra and temporal characteristics are measured using optical spectrum analyzers (OSA~1 and OSA~2), a fast photodetector (PD) with oscilloscope (OSC), and an intensity autocorrelator (AC). ISO: isolator; CIR: circulator; EDFA: erbium-doped fiber amplifier
		(b) Photograph of the integrated MLL in operation.
		(c) Stable pulse train acquired via direct photodetection of the MLL output.
		(d) Optical intensity autocorrelation traces after optimized linear chirp compression using either a programmable filter (pulse shaper) and EDFA (Path A), or through $\sim$10\,m of single-mode silica fiber (SMF) (Path B).
		(e) Evolution of the autocorrelation function under varying group delay dispersion compensation.
		(f) Optical spectra acquired at the two output ports of the integrated MLL at high pump power.
		(g) Spectrogram of output 1 recorded continuously over 10~hours when pumped with 820~mW pump 2.
	}
	\label{fig2}
\end{figure*}

Figure\,\ref{fig1}(a) illustrates the working principle of the integrated MLL based on a Mamyshev oscillator.
A 42-cm-long section of \ErSiN{} waveguide is placed between two spectrally offset WBGs, forming a simple linear cavity with 175.5~MHz free spectral range.
As simulated in Fig.\,\ref{fig1}(c) (see Supplementary Information S2), pulses reflected by one grating undergoes simultaneous amplification and self-phase modulation, broadening its spectrum and bridging the spectral gap between the WBGs, eventually establishing stable mode-locked lasing.

To maximize nonlinear interactions while maintaining low optical loss (see Supplementary Information S3), we employ a narrow few-mode \SiN{} waveguide with a cross-section of $1.6\,\mathrm{\mu m} \times 0.35\,\mathrm{\mu m}$ (Fig.\,\ref{fig1}(e)).
The devices were fabricated using a 4-inch wafer-scale process, as detailed in Methods.
Erbium ions were implanted into the waveguide to provide optical gain~\cite{liu2021high}, using a maximum implantation energy of 500~keV and a total fluence of $3.87\times10^{15}\,{\rm cm^{-2}}$.
This results in an estimated peak doping concentration of approximately 0.4 atomic percent and a 22\% overlap between the mode and the dopant profile~\cite{liu2021high} (Fig.\,\ref{fig1}(f)).
The WBGs, shown in Fig.\,\ref{fig1}(g), serve as narrowband reflectors for the pulses, while simultaneously enabling injection of the 1480~nm~pump and partial transmission for MLL output.
We adopt an apodized grating design~\cite{erdogan2002fiber}, using only 950 grating periods to reduce both chromatic dispersion and cladding mode scattering loss at the pump wavelength~\cite{gratingscatteringloss}, while maintaining sufficient reflectivity.
The fabricated WBGs exhibit 3~dB bandwidths of approximately 5~nm and a center wavelength separation of 8.9~nm (Fig.\,\ref{fig1}(h)).
The reflection bands are designed near 1557~nm to optimally leverage the erbium gain bandwidth.
Microheaters are placed above the WBGs to facilitate fine-tuning of their reflection wavelengths (Fig.\,\ref{fig1}(b)).
For prototyping, 26 independent MLLs were placed in a parallel waveguide bundle and collectively routed in a spiral layout on a $2~\mathrm{cm}\times 1.1~\mathrm{cm}$ chip (Fig.\,\ref{fig1}(d)), yielding over 300 MLLs per wafer.

\section{Mode-locking operation and characterization}
Figure\,\ref{fig2}(a) illustrates the experimental setup used to operate and characterize the integrated MLL (see Supplemental Information S4 for details).
For maximum output power, the MLL was optically pumped from both ends of the waveguide with two unpolarized 1480~nm lasers (Pump 1 and Pump 2), delivering 959~mW and 820~mW off-chip optical power, respectively.
These power levels are readily achievable using commercially available single-transverse-mode indium phosphide laser diodes after polarization combining.
Similar to fiber-based Mamyshev oscillators~\cite{regelskis2015ytterbium,liu2017megawatt}, the integrated laser was seeded with a single pulse from an external MLL.
The seed pulse was filtered to limit the 3~dB bandwidth to 0.15~THz and gated using an electro-optic intensity modulator. 
Upon seeding, self-sustained mode-locking was established without continuous-wave background or Q-switching instabilities, as confirmed by the stable pulse train detected using a DC-coupled fast photodetector (Fig.\,\ref{fig2}(c)).
While external seeding was used in the demonstrations, we have also shown that mode-locking can be initiated by generating a Q-switched pulse within the cavity using an external reflector and modulator, eliminating the need for a seed source (see Supplemental Information S4 and S5).
The optical spectra collected from both waveguide outputs (Fig.\,\ref{fig2}(f)) centered at approximately 1.56\,$\mu$m and showed 64~nm and 47~nm of 20~dB bandwidth, spanning the entire telecom C-band and extending into the L-band.
This corresponds to over 40,000 comb lines (not resolved in the spectrum).
We measured average output powers of 136~mW (output 1) and 138~mW (output 2) from the two ends of the linear cavity, corresponding to on-chip powers of 182~mW and 184~mW after accounting for coupling loss (see Supplementary Information S6), yielding a pump-to-output efficiency of 27.5\%.
With a repetition rate $f_\mathrm{rep}$ of 175.5~MHz, the pulse energies at each output reached 1.04~nJ and 1.05~nJ — over two orders of magnitude higher than previously reported photonic integrated ultrafast sources.
For power-constrained applications, stable mode-locking was also achieved using only the 820~mW pump power delivered by Pump 2.
Under this lower power condition, the oscillator produced on-chip pulse energies of 397~pJ and 347~pJ from outputs 1 and 2, respectively, still surpassing all prior integrated ultrafast sources by a wide margin.
As shown by the spectrogram (Fig. \ref{fig2}(g)), the integrated MLL maintained stable mode-locking for over 10 hours, demonstrating the robustness of the mode-locked state and the long-term integrity of the waveguide.
A photograph of the device in operation is shown in Fig.\,\ref{fig2}(b).

The output pulses from the integrated Mamyshev oscillator were predominantly linearly chirped, in agreement with numerical simulations (see Supplemental Information S2).
To characterize the pulses, we used an intensity autocorrelator in conjunction with a programmable filter to apply a variable group delay dispersion, and a home-built EDFA to compensate for loss in the filter.
Sweeping the applied dispersion, the pulse undergoes compression and re-broadening (Fig.\,\ref{fig2}(e)), allowing identification of the optimal dispersion for pulse compression.
Moreover, the pulse waveform can be computationally reconstructed from the dispersion sweep data (see Supplementary Information S7). 
With the optimized group delay dispersion, the intensity autocorrelation function of the output pulse from output 1 was compressed to 187~fs full-width half maximum (Fig.\,\ref{fig2}(d)).
Alternatively, using a SMF delay line ($\sim10~\mathrm{m}$ long including other components in the setup) for dispersion compensation, the pulse from output 2 was directly compressed to a 147~fs autocorrelation width.
Such dispersion compensation can be readily implemented on \SiN{} platform with chirped Bragg gratings~\cite{du2020silicon}.

\section{Coherence of the integrated mode-locked laser}
\begin{figure*}
	\centering
	\includegraphics[width=1\linewidth]{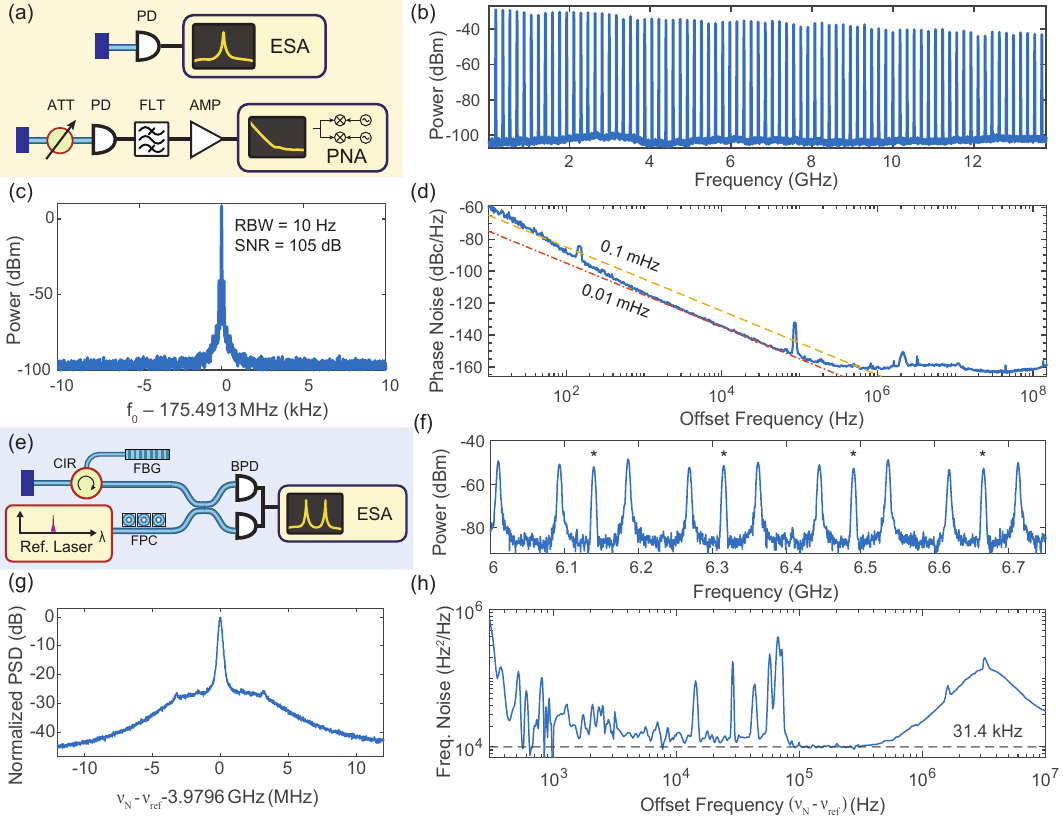}
	\caption{\textbf{Coherence characterization of the integrated mode-locked laser.}
		(a) Schematic for repetition-rate beatnote ($f_{\rm rep}$) characterization via direct photodetection.
		ATT: attenuator; FLT: filter; AMP: amplifier.
		(b) Spectrum of the photodetected signal, showing multiple harmonics of the repetition rate\,$f_{\rm rep}$.
		(c) Measured\,$f_{\rm rep}$ beatnote with 105~dB signal-to-noise ratio (SNR) at 10~Hz resolution bandwidth (RBW).
		(d) Phase noise of the\,$f_{\rm rep}$ beatnote, with reference lines corresponding to 0.1~mHz and 0.01~mHz Lorentzian linewidth.
		(e) Schematic for heterodyne beatnote characterization of the integrated MLL, where a portion of the MLL spectrum is preselected by a fiber Bragg grating (FBG) and interferences with a reference continuous-wave laser.
		(f) Spectrum of the heterodyne beatnote between the reference laser and the integrated MLL.
		Asterisks denote residual\,$f_{\rm rep}$ harmonics resulting from imperfect common-mode suppression in the balanced photodetector (BPD).
		(g) Power spectral density (PSD) of the heterodyne beatnote centered at 3.9796~GHz with a RBW of 12.47~kHz.
		The linewidth for a 10~ms measurement time is 174~kHz.
		(h) Frequency noise of the heterodyne beatnote in (g), indicating a 31.4~kHz Lorentzian linewidth for the optical comb line of the integrated MLL.}
	\label{fig3}
\end{figure*}

For many applications, MLLs are desired to be low noise in terms of pulse interval and pulse-to-pulse phase slip~\cite{Kim16}. 
We evaluated the coherence of the integrated Mamyshev oscillator via direct photodetection (Fig.\,\ref{fig3}(a)) and heterodyne beating with a reference laser (Fig.\,\ref{fig3}(e)).
The photodetected signal exhibits tens of harmonics of repetition-rate $f_\mathrm{rep}$, extending well into the microwave range (Fig.\,\ref{fig3}(b)).
Using a low-noise amplifier and appropriate filtering, the signal-to-noise ratio (SNR) of the fundamental beatnote at $f_\mathrm{rep}$ was measured to be $105~\mathrm{dB}$ at $10~\mathrm{Hz}$ resolution bandwidth, approaching the limit of the instrument.
To estimate the timing jitter of the MLL, the single-sideband (SSB) phase noise power spectral density (PSD) of the sixth repetition-rate harmonic was characterized with a phase noise analyzer (PNA).
After scaling down to the fundamental $175.5~\mathrm{MHz}$ beatnote, the SSB phase noise at $10~\mathrm{kHz}$ offset reaches $-134~\mathrm{dBc/Hz}$.
The integrated timing jitter between $10~\mathrm{kHz}$ and $10~\mathrm{MHz}$ is $59.1~\mathrm{fs}$.
Moreover, by fitting the $1/f^2$ slope to the $f_\mathrm{rep}$ phase noise between $1~\mathrm{kHz}$ and $50~\mathrm{kHz}$ offset, we extracted a Lorentzian linewidth of $0.012~\mathrm{mHz}$, which represents a significant improvement over the best reported value from heterogeneously integrated semiconductor MLLs~\cite{lownoisehetero2021}.

The heterodyne beatnote between the MLL and a continuous-wave reference laser at 1552~nm (Fig.\,\ref{fig3}(f)) reveals a series of regularly spaced tones, indicating a stable frequency comb with a well-defined carrier-envelope offset frequency over the measurement duration.
To quantify the linewidth of individual comb lines (Fig.\,\ref{fig4}(g)), we acquired the downconverted signal with an electronic signal analyzer (ESA) and extracted its SSB frequency noise PSD (Fig.\,\ref{fig3}(h)).
By fitting the noise floor near $200~\mathrm{kHz}$ offset, we determined a Lorentzian linewidth of 31.4~kHz.
Spurs observed between $10~\mathrm{kHz}$ and $100~\mathrm{kHz}$ are likely caused by the intensity noise of the 1480~nm pump laser, while the bump between $1~\mathrm{MHz}$ and $10~\mathrm{MHz}$ may be attributed to relaxation oscillation.
The frequency noise between 1~kHz and 100~kHz offset remained below $5\times 10^4~\mathrm{Hz^2/Hz}$, which we attribute to the large mode volume of the cavity and thus low thermorefractive noise~\cite{guanhaoTRN}.

\section{Direct supercontinuum generation in \SiN{} waveguide}
\begin{figure*}
	\centering
	\includegraphics[width=1\linewidth]{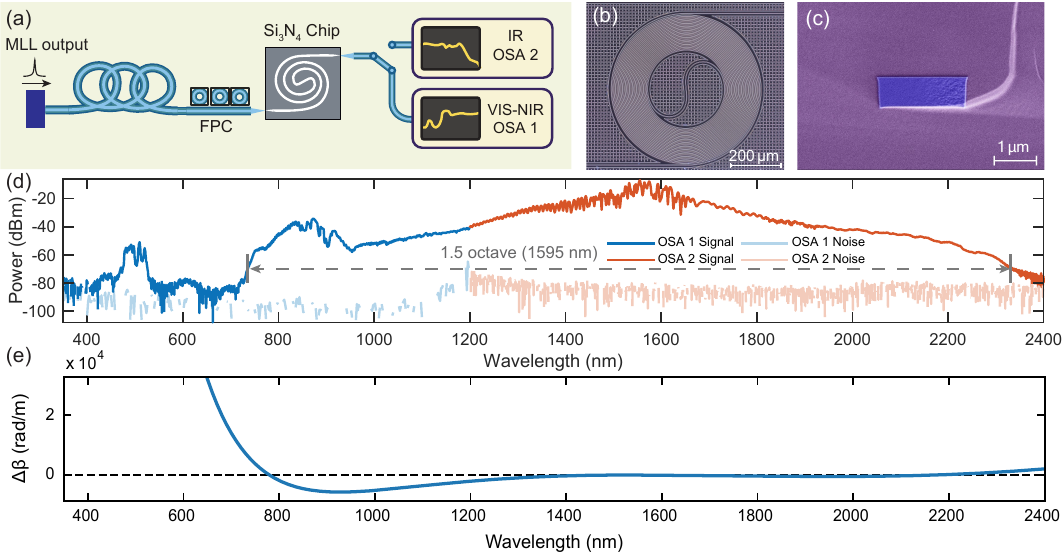}
	\caption{\textbf{Supercontinuum generation directly driven by the integrated Mamyshev oscillator.}
		(a) Experiment setup for supercontinuum generation in a \SiN{} spiral waveguide, directly driven by the integrated MLL without post-amplification.
		(b) Photograph of a \SiN{} spiral waveguide chip identical to the one used for supercontinuum generation.
		The internal tracking number of the sample used in the experiment is \texttt{D7802F4C4}.
		(c) False-colored SEM image of the waveguide cross-section.
		(d) Spectrum of the generated 1.5-octave spanning supercontinuum.
		The -20 dB bandwidth is 18.7~THz.
		(e) Simulated integrated dispersion of the spiral waveguide used in the experiment.
	}
	\label{fig4}
\end{figure*}
A compelling application of high-energy on-chip pulse sources is the realization of fully integrated, self-referenced optical frequency combs, which require the generation of a broad optical spectrum.
As a proof of concept, we routed the output of the integrated MLL through approximately 10~m of SMF, followed by coupling into a 43.7~mm \SiN{} spiral waveguide on a separate chip for supercontinuum generation (Fig.,\ref{fig4}(a)).
This fiber link also provides dispersion compensation for the chirped MLL pulses, compressing them to reach kilowatt-level peak power.
The \SiN{} spiral waveguide (Fig.\,\ref{fig4}(b)), fabricated using the photonic damascene process~\cite{liu2021high}, features a $2.07\,\mathrm{\mu m} \times 0.70\,\mathrm{\mu m}$ cross-section (Fig.\,\ref{fig4}(c)).
It exhibits a relatively flat integrated dispersion near telecom C-band and a dispersive wave phase matching in the near-visible wavelength (Fig.\,\ref{fig4}(e)).
Without optical amplification, the MLL output drives a 1.5-octave-spanning supercontinuum from 736~nm to 2331~nm (Fig.\,\ref{fig4}(d)), corresponding to a 279~THz bandwidth and is sufficient for $f$–$2f$ self-referencing. 
The measured spectrum matches with numerical simulation (see Supplementary Information S8), while the observed signal below 600~nm may result from modal-phase-matched third-harmonic generation.

\section{Discussion and conclusions}

We have demonstrated a fully integrated Mamyshev oscillator architecture MLL based on a wafer-scale fabricated erbium-doped \SiN{} photonic chip, capable of generating ultrafast pulses with nanojoule-level energy. 
The laser delivers pulses that can be compressed to 147~fs and exhibits excellent coherence compared to existing integrated MLLs.
Coherence may be further enhanced by using lower-noise pump sources, lowering waveguide loss by fabrication optimizations (see Supplementary Information S9) and optimizing cavity dispersion.
The nanojoule-level pulses make a range of on-chip nonlinear processes directly accessible.
As a proof of concept, we demonstrated 1.5-octave-spanning supercontinuum generation in a separate \SiN{} waveguide, without any optical amplification.
In the future, this supercontinuum generation stage can be seamlessly co-integrated with the MLL chip.
Currently, the MLL requires an external pulse source or an extended Q-switched arm to initiate mode-locking.
Self-seeding of the integrated Mamyshev oscillator may be achieved by fast modulation of the WBGs~\cite{wang2023automated}, for example via the Pockels effect in heterogeneously integrated lithium niobate, which can generate similar Q-switched pulses.
While our demonstration is based on silicon nitride waveguides and in the 1.55~$\mathrm{\mu m}$ band, the Mamyshev oscillator concept is readily transferable to other platforms and wavelengths. 
The combination of high pulse energy, excellent coherence, and stable mode-locking demonstrated here opens new opportunities for nonlinear integrated photonics, including mid-infrared supercontinuum sources for spectroscopy~\cite{guo2018mid}, compact terahertz systems for non-destructive testing~\cite{tonouchi2007cutting}, and chip-scale frequency combs for optical atomic clocks~\cite{JunyeClockRevModPhys}.

\clearpage

\section*{Methods}

\begin{footnotesize}

\linespread{0}%
\noindent\textbf{Sample fabrication process} 
The erbium-doped wafers used in this study (\verb|D20603|) were fabricated using a subtractive process~\cite{ji2024efficient,ji2025full}.
The fabrication starts from commercial 4-inch silicon wafers (SIEGERT WAFER GmbH) with a \SI{10}{\micro\metre} thermal \ce{SiO2} layer grown by wet oxidation.
A \SI{362}{\nano\metre} stoichiometric \ce{Si3N4} film was then deposited via low-pressure chemical vapor deposition (LPCVD).
The slightly increased deposition thickness compensates for shrinkage during subsequent annealing, resulting in a final thickness close to \SI{350}{\nano\metre}.
The waveguide pattern was defined using deep ultraviolet (DUV) lithography (ASML PAS5500/350C, JSR M108Y, Brewer Science DUV-42P), followed by reactive ion etching (RIE) with a fluorine-based chemistry (\ce{CHF3}, \ce{SF6}, \ce{Ar}, \ce{O2}) in a SPTS APS system.
The \ce{Si3N4} etching recipe was optimized for vertical sidewalls.
However, the critical dimension of isolated features shrank by approximately \SI{55}{\nano\metre} on each side, which was compensated for in the design of the edge couplers and calibration of the waveguide Bragg grating parameters.
After resist stripping, the backside \ce{Si3N4} (from double-sided LPCVD) was removed by RIE, while protecting the front side with photoresist.
Megasonic cleaning was performed to remove particle contamination on the wafer after etching with front side facing down.
The wafers were then annealed at \SI{1200}{\celsius} for \SI{12}{\hour} in \ce{N2}.
Removing the backside \ce{Si3N4} prior to annealing is crucial, as the \SI{350}{\nano\metre} film has sufficient internal stress to cause plastic deformation and permanent bowing of the wafer (up to \SI{170}{\micro\metre}), which would impede subsequent fabrication steps.
The wafers then underwent standard ultraviolet (UV) direct-write lithography (Heidelberg MLA150, \SI{6}{\micro\metre} AZ 15nXT) to define the mask for ion implantation, protecting the waveguide Bragg gratings and edge couplers from damage due to implantion.

The erbium ion implantation was performed at the Ion Beam Center of HZDR using a 500~kV air-insulated implanter.
Singly charged ions were extracted from an indirectly heated cathode Bernas-type ion source with a 40~kV extraction system, and post-accelerated to energies up to 500~keV.
The ions were then mass-separated according to their charge-to-mass ratio using an analyzing magnet.
The beam was guided through the line using quadrupoles and an Einzel lens.
A neutral trap was employed to prevent neutral particles from reaching the target.
To ensure uniform irradiation, the ion beam was scanned over the sample at approximately 1~kHz in both the X and Y directions, with slightly offset frequencies to improve spatial homogeneity.
The ion implantation was performed at three different energies consequently to tailor a dopant profile across the film thickness.
The implantation was carried out at three energies: 125~keV, 253~keV, and 500~keV and the corresponding implantation doses were $8.40 \times 10^{14}$~cm$^{-2}$, $1.41 \times 10^{15}$~cm$^{-2}$, and $2.84 \times 10^{15}$~cm$^{-2}$, respectively. 
All implantations were conducted at room temperature with a $0^\circ$ incidence angle.

After the ion implantation, the photoresist was removed by oxygen plasma.
The surface was then cleaned with concentrated \ce{HCl} solution to eliminate erbium oxide residue and diluted \ce{HF} solution to remove resputtered material.
The wafers were subsequently annealed at \SI{900}{\celsius} for \SI{1}{\hour} in \ce{N2} to repair radiation damage to the \ce{Si3N4} waveguide.
Approximately \SI{6}{\micro\metre} of \ce{SiO2} cladding was deposited using inductive coupled plasma chemical vapor deposition (ICPCVD) with a \ce{SiCl4} precursor~\cite{qiu2023hydrogen}.
The wafers were then annealed at \SI{600}{\celsius} for \SI{1}{\hour} in \ce{O2} to reduce the optical loss caused by the cladding process.
Metallic microheaters were fabricated by sputtering $\sim$\SI{25}{\nano\metre} of \ce{Ti} and $\sim$\SI{500}{\nano\metre} of \ce{Pt}, followed by UV lithography (Heidelberg MLA150, \SI{2}{\micro\meter} AZ 10nXT) and ion beam etching (Veeco IBE350).
The photoresist was reflown at \SI{135}{\celsius} to reduce resputtering caused fencing during etching.
Die separation was carried out using UV lithography (Heidelberg MLA150, \SI{10}{\micro\meter} AZ 15nXT), followed by standard deep RIE of the \ce{SiO2} layer and approximately \SI{250}{\micro\metre} of \ce{Si}, and concluding with backside grinding to thin down the chip to around \SI{250}{\micro\metre}.
The finished chips were baked on a \SI{310}{\celsius} hot plate overnight to heal \ce{Si3N4} damage incurred from UV exposure caused by emission from intense plasma throughout the fabrication process, which can increase the optical loss of waveguides~\cite{ji2024efficient}.

The dispersion-engineered \ce{Si3N4} samples used for the supercontinuum generation demonstration (\texttt{D7803}) were fabricated using the photonic damascene process, as described in~\cite{liu2021high}.

\linespread{0}%
\noindent \textbf{Author Contributions}: 
Z.Q., Z.S.L. and J.H. conceived the work.
Z.Q. and J.H. performed numerical simulation and designed the \texttt{D206} devices, with help from J.S. and Z.S.L..
Z.Q. fabricated the device (excluding ion implantation) with substantial help from Y.Z., X.J., X.L., and Z.H.L.
U.K. performed the ion implantation.
Z.Q., X.Y. and J.H. performed the experiment and processed the data, with help from G.L. and X.L..
Z.Q. and J.H. wrote the manuscript with contributions from all coauthors.
T.J.K. supervised the work.\\
\noindent \textbf{Funding Information and Disclaimer}:
This work was supported by funding from the Swiss National Science Foundation under grant agreement No.216493 (HEROIC).
Z.Q., X.Y., and T.J.K. are co-inventors on European patent application EP25200343.9 concerning the integrated Mamyshev oscillator.\\
\noindent \textbf{Acknowledgments}:
The samples were partially fabricated in the EPFL Center of MicroNanoTechnology (CMi).
The ion implantation was carried out at Helmholtz-Zentrum Dresden-Rossendorf (HZDR).
We thank Johann Riemensberger for designing and Junqiu Liu and Rui Ning Wang for fabricating the \texttt{D7803} sample used for supercontinuum generation demonstration.\\
\noindent \textbf{Data Availability Statement}: The code and data used in this work will be provided in a \texttt{Zenodo} repository upon publication of this preprint.
\par
\end{footnotesize}
\bibliography{ref}
	
\end{document}


\title{Supplementary information for:\\High-pulse-energy integrated mode-locked lasers based on a Mamyshev oscillator}

\newcommand{\epflaff}{\affiliation{\mbox{Institute of Physics}, \mbox{Swiss Federal Institute of Technology Lausanne (EPFL)}, CH-1015 Lausanne, Switzerland}
\affiliation{\mbox{Institute of Electrical and Microengineering}, EPFL, CH-1015 Lausanne, Switzerland}}
\newcommand{\hzdraff}{\affiliation{Helmholtz-Zentrum Dresden-Rossendorf (HZDR), 01328 Dresden, Germany}}
\newcommand{\cofirst}{\textsuperscript{\textdagger}}

\newcommand\blfootnote[1]{
  \begingroup
  \renewcommand\thefootnote{}\footnote{#1}%
  \addtocounter{footnote}{-1}%
  \endgroup
}
\author{Zheru Qiu\cofirst}\epflaff
\author{Jianqi Hu\cofirst}\epflaff
\author{Xuan Yang\cofirst}\epflaff
\author{Zhongshu Liu}\epflaff
\author{Yichi Zhang}\epflaff
\author{Xinru Ji}\epflaff
\author{Jiale Sun}\epflaff
\author{Grigory Lihachev}\epflaff
\author{Xurong Li}\epflaff
\author{Zihan Li}\epflaff
\author{Ulrich Kentsch}\hzdraff
\author{Tobias J. Kippenberg$^\ast$}\epflaff
\medskip
\maketitle
\blfootnote{\cofirst:These authors contributed equally.\\$^\ast$: \href{mailto:tobias.kippenberg@epfl.ch}{tobias.kippenberg@epfl.ch}}
\setcounter{equation}{0}
\setcounter{figure}{0}
\setcounter{table}{0}
\setcounter{subsection}{0}
\setcounter{section}{0}
\tableofcontents
\setcounter{page}{1}
\newpage
\setcounter{secnumdepth}{3}
\renewcommand\thepage{S\arabic{page}}

\section{Comparison with state-of-the-art integrated ultrafast sources}
\begin{figure*}[htbp]
	\centering
	\includegraphics[width=0.8\linewidth]{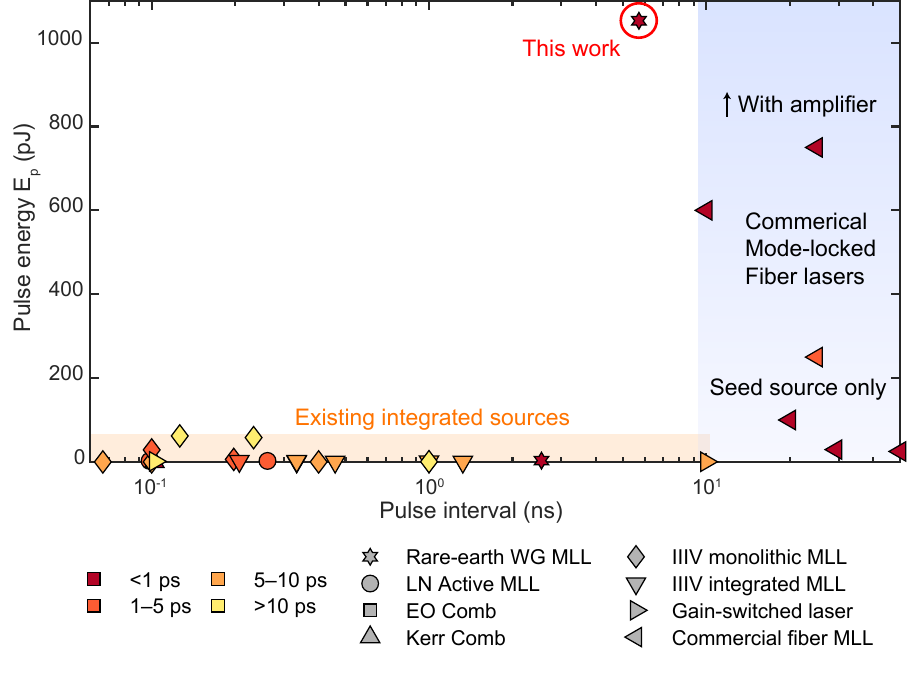}
	\caption{Comparison with state-of-the-art integrated ultrafast pulse sources in terms of pulse energy $E_\mathrm{p}$ and pulse interval ($1/f_\mathrm{rep})$.
		The color of the markers represent the pulse width.
	}
	\label{figS:compare}
\end{figure*}

Here, we compare the key performance metrics of our work with state-of-the-art integrated ultrafast sources reported in the literature (Table~\ref{tab:ultrafast_sources}).
For completeness, we present a comprehensive list of integrated ultrafast sources, including some that may not strictly be considered photonic integrated circuits (PICs) by today’s standards.
In addition, a selection of commercially available low-power table-top mode-locked lasers (MLLs) is also listed for reference.
Figure~\ref{figS:compare} compares the pulse energy $E_{\mathrm p}$ versus the pulse interval ($1/f_{\mathrm{rep}}$) of the sources listed in Table~\ref{tab:ultrafast_sources}.

\FloatBarrier
\setlength{\LTcapwidth}{6in}
\begin{ThreePartTable}
	\renewcommand\TPTminimum{\textwidth}
	\setlength\LTleft{0pt}
	\setlength\LTright{0pt}
	\setlength\tabcolsep{0pt}
	
	\newcolumntype{P}[1]{>{\arraybackslash}p{#1}}
	\begin{longtable}{%
			P{0.8cm}                        
			P{0.88cm}                        
			P{0.97cm}                          
			P{1cm}                        
			P{1.35cm}                        
			P{1cm}                        
			P{1.6cm}                        
			P{1.6cm}                        
			P{2.2cm}                        
			P{4.1cm}                        
			r                                
		}
		
		\caption{Comparison with state-of-the-art integrated ultrafast pulse sources.}
		\label{tab:ultrafast_sources} \\  
		\toprule
		\makecell{Year} &
		\makecell{$\lambda_c$ \\ ($\mu$m)} &
		\makecell{$f_{\mathrm rep}$ \\ (GHz)} &
		\makecell{Pulse \\ width \\ (ps)} &
		\makecell{On-chip \\average\\  power \\ (mW)} &
		\makecell{$E_{\mathrm p}$ \\ (pJ)} &
		\makecell{$f_{\mathrm{rep}}$ RF\\ linewidth \\ (Hz)} &
		\makecell{Comb line \\ linewidth \\ (kHz)} &
		\makecell{Spectrum\\width \\ (nm)} &
		\makecell{Type} &
		Reference \\
		\midrule
		\endfirsthead
		
		\multicolumn{4}{l}{\small\itshape Table~\thetable\ (continued)}\\
		\toprule
		\makecell{Year} &
		\makecell{$\lambda_c$ \\ ($\mu$m)} &
		\makecell{$f_{\mathrm rep}$ \\ (GHz)} &
		\makecell{Pulse \\ width \\ (ps)} &
		\makecell{On-chip \\average\\  power \\ (mW)} &
		\makecell{$E_{\mathrm p}$ \\ (pJ)} &
		\makecell{$f_{\mathrm{rep}}$ RF\\ linewidth \\ (Hz)} &
		\makecell{Comb line \\ linewidth \\ (kHz)} &
		\makecell{Spectrum\\width \\ (nm)} &
		\makecell{Type} &
		Reference \\
		\midrule
		\endhead
		
		\bottomrule
		\endlastfoot
		
		2025 & 1.56 & 0.175 & 0.147$^a$ & 184 & 1052 & $1.2\times 10^{-5}$ & 31.4 & \makecell{64 (20 dB)} & Er-doped-\ce{Si3N4} Mamyshev & \makecell{\bf This work} \\
		\midrule
		2023 & 1.06 & 10.17 & 4.81 & 53 & 2.6 & x & $3.91\times10^3$ & \makecell{0.35 (3 dB)} & LN active MLL & \cite{guo2023ultrafast} \\
		1995 & 1.60 & 3.823 & 3.8 & 0.9 & 2.4 & x & $-$ & \makecell{0.95 (3 dB)} & LN active MLL & \cite{suche1995harmonically} \\
		2024 & 1.55 & 39.58 & $\sim$20 & 11 & 0.28 & x & 0.6${}^\mathrm{v}$ & 20 & LN active MLL & \cite{ling2024electrically} \\
        2009 & 1.56 & 0.393 & 0.44 & 1.2 & 3.05 & + & - & 8.4 (3 dB) &  Er-glass WG MLL (SESAM) & \cite{byun2009integrated,Pudo08} \\
		2019 & 1.90 & 1.2 & - & 9 & - & - & - & 17 (3 dB) & \ce{Tm\dop Al2O3} passive MLL & \cite{shtyrkova2019integrated} \\
		2022 & 1.56 & 30.10 & 0.52 & 16.25 & 0.54 & x & x & \makecell{6 (10 dB)} & EO comb (LN, time-lens) & \cite{yu2022integrated} \\
		2022 & 1.55 & 30.925 & 0.336 & 20 & 0.65 & x & x & 132 ($-70~\mathrm{dBm}$) & EO comb (LN, resonant) & \cite{hu2022high} \\
		2020 & 1.56 & 9.78 & 0.098 & 0.11$^c$ & 0.011$^c$ & x & x & 25.8 (3 dB) & Kerr comb (SiN) & \cite{2020NaPho..14..486L} \\
		2023 & 1.55 & 99.72 & ++ & $\sim$6 & $\sim$0.06 & x & x & $-$ & Kerr comb (SiN) & \cite{helgason2023surpassing} \\
		2022 & 1.55 & 143.6 & ++ & 20 & 0.14 & x & 0.5 & 200 (60 dB) & Kerr comb (hybrid SiN) & \cite{dmitriev2022hybrid}\\
		2006 & 1.55 & 4.29 & 10 & 250 & 58 & - & - & 5.7 (3 dB) & III--V monolithic (SCOWL)& \cite{plant2006250} \\
		2007 & 0.98 & 7.92 & 16 & 489 & 62 & - & - & 0.96 (3 dB) & III--V monolithic (SCOWL)& \cite{4214866} \\
		2005 & 1.26 & 5.06 & 3.2 & $<$30 & $<$6 & - & - & 7.1 (3 dB) & III--V monolithic (QD)& \cite{gubenko2005high} \\
		2011 & 1.51 & 10.0 & 6 & 0.8$^o$ & 0.08$^o$ & $1\times10^2$$^e$ & - & 1.5 & III--V monolithic (QD)& \cite{carpintero2011comparison} \\
		2005 & 1.26 & 21.0 & 0.39 & 25 & 0.95 & - & - & 20 & III--V monolithic (QD)& \cite{Rafailov} \\
        2008 & 1.54 & 92.0 & 0.312 & 8.6 & 0.09 & - & - & 11.62 (3 dB) & III--V monolithic (QD) & \cite{lu2008312} \\
		2012 & 1.26 & 10.0 & 2.2 & 288 & 28.7 & - & - & 5.6 (3 dB) & III--V monolithic (Tapered) & \cite{nikitichev2012high} \\
		2025 & 1.53 & 10.6 & 0.726 & - & - & - & - & 19.22 (20 dB) & III--V monolithic (QW) & \cite{SUN2025131655} \\
		2006 & 1.53 & 15.0 & 5 & $\sim$0.15 & 0.01 & $4\times10^5$ & $6\times10^5$ & 4.5 (3 dB) & III--V monolithic (Ring) & \cite{barbarin2006characterization} \\
		2015 & 1.58 & 2.5 & 9.8 & 0.08 & 0.032 & $6.13\times 10^3$ & - & 3 (3 dB) & III--V monolithic (InP PIC)& \cite{Latkowski15} \\
		2017 & 1.53 & 21.5 & 0.35 & 1 & 0.047 & $4.5\times10^5$ & - & 14 (3 dB) & III--V monolithic (InP PIC)& \cite{lo20171} \\
		2010 & 1.58 & 1.0 & 36 & $>$0.59 & 0.59 & - & $7\times10^4$ & 5 (10 dB) & III--V monolithic (QW)& \cite{5598511} \\
		2021 & 1.55 & 0.99 & 272 & - & - & $7\times10^4$ & - & - & III--V monolithic & \cite{9439481} \\
		2019 & 1.31 & 20.0 & 1.7 & 0.48 & 0.024 & 400 & - & 5.5 (3 dB) & III--V monolithic (on Si) & \cite{Auth:19} \\
		2019 & 1.27 & 20.0 & 5 & $<$18 & $<$0.9 & $1.8\times10^3$ & $10.6\times10^3{}^\mathrm{v}$ & 8.4 (10 dB) & III--V monolithic (on Si) & \cite{liu2019high} \\
		2024 & 1.55 & 3.0 & - & 2 & 0.67 & - & - & 23 (10 dB) & III--V integrated (het. SiN) & \cite{Billet24} \\
		2018 & 1.58 & 20.0 & 0.9 & 1.83 & 0.09 & $1.1\times10^3$ & - & $\sim$20 (20 dB) & III--V integrated (het. Si) & \cite{Davenport:18} \\
		2021 & 1.61 & 3.0 & 8 & 5.6 & 2 & $4\times10^2$ & $<10^3$ & 4 (10 dB) & III--V integrated (het. SiN) & \cite{Hermans21} \\
		2015 & 1.56 & 4.83 & 3 & 9 & 1.86 & $1.7\times10^3$ & $<2\times10^3$ & 3.5 (10 dB) & III--V integrated (het. Si) & \cite{keyvaninia2015iii} \\
		2021 & 1.58 & 0.755 & 7.46 & 0.125 & 0.17 & 1 & 146 & 3.27 (10 dB) & III--V integrated (het. SiN) & \cite{lownoisehetero2021} \\
		2017 & 1.60 & 1.0 & 7 & $<$0.8 & $<$0.8 & $9\times10^2$ & $2.5\times10^2$ & 17 (10 dB) & III--V integrated (het. Si) & \cite{wang2017iii} \\
		2021 & 1.56 & 2.18 & 6.31 & 0.4$^o$ & 0.18 & 31 & - & 8.3 (10 dB) & III--V integrated (hyb. SiN) & \cite{Vissers21} \\
		2023 & 1.27 & 0.1 & 5.3 & 0.042 & 0.42 & - & - & 0.4 (3 dB) & Gain-switched DFB laser & \cite{Kobayashi:23} \\
		2021 & 1.55 & 9.72 & - & 13 & 1.34 & - & - & - & Gain-switched laser & \cite{weng2021gain} \\
		\midrule
		2025 & 1.55 & 0.04 & $<$0.3 & 30 & 750 & - & - & 20 & Fiber MLL (with amp.) & \cite{OptilabFML15PM} \\
		2025 & 1.56 & 0.1 & 0.1 & 60 & 600 & $3\times10^{-8}$$^{\text{p}}$ & $<0.12^{\text{p}}$ & 30 & Fiber MLL (with amp.) & \cite{MenloELMO2025} \\
		2025 & 1.50 & 0.034 & 0.38 & 1 & 29.4 & - & - & 20 & Fiber MLL (without amp.) & \cite{SymphotonyFLMLEr2025} \\
		2025 & 1.95 & 0.05 & 0.35 & 5 & 100 & - & - & 10 & Fiber MLL (without amp.)& \cite{AdValueAPML2025} \\
		2025 & 1.03 & 0.04 & 3 & 10 & 250 & $3\times10^{-10}$$^\text{e}$ & - & 14 & Fiber MLL (without amp.) & \cite{CycleSonata2025} \\
		2025 & 1.55 & 0.02 & 0.5 & 0.5 & 25 & - & - & - & Fiber MLL (without amp.) & \cite{Calmar1550LowJitter2020}
	\end{longtable}
	\begin{tablenotes}
        \item \hspace{-1.2em}{\small \textit{Notes:}}
		\footnotesize
        \item \hspace{-1.2em} QD: Quantum dot; QW: Quantum well; SCOWL: Slab-coupled optical waveguide laser; SESAM: Semiconductor saturable absorber mirrors; LN:\ce{LiNbO3}; het.: heterogeneous; hyb.: hybrid.
		\item \hspace{-1.2em} x Not applicable (for example, directly depend on the driving source).
		\item \hspace{-1.2em} - No data provided.
		\item \hspace{-1.2em} + No data provided, but likely good as the integrated jitter (26~fs from 10~kHz to 10~MHz) is low.
		\item \hspace{-1.2em} ++ No data provided, but likely very short due to the solitonic nature and wide spectrum.
		\item[a] Autocorrelation function width after linear compression in fiber.
		\item[c] Estimated from the conversion efficiency. 
		\item[p] Experimentally measured at EPFL for an older unit produced in 2016.
		\item[e] Estimated from figures on datasheet.
		\item[o] Off-chip coupled power. No coupling efficiency reported.
        \item[v] Average optical linewidth of each mode from delayed self-heterodyne method.
	\end{tablenotes}
\end{ThreePartTable}

\newcommand{\fntxt}{%
	In some literature on erbium-doped fibers or waveguides, $P_\mathrm{sat}$ is often presented as a constant independent of pump power, and typically expressed as inversely proportional to the upper-level lifetime $\tau$ and the emission (or absorption) cross-section $\sigma$.
	However, this is incorrect for systems where the optical pumping rate greatly exceeds the spontaneous decay rate, as can be seen by analyzing the rate equations Eq. \eqref{eq:rate}.
	For erbium-doped waveguides, while the spontaneous decay rate is on the order of $10^2~\mathrm{s^{-1}}$ due to the millisecond-scale lifetime, the optical pumping rate can reach $10^5~\mathrm{s^{-1}}$ for hundreds of milliwatts of pump power.
	When the waveguide functions as a laser cavity with signal average power exceeding several milliwatts (as in continuous-wave or high-repetition-rate pulsed systems), the stimulated emission rate can similarly exceed the spontaneous decay rate by orders of magnitude, making $\tau$ irrelevant.%
}
\section{Simulation of integrated Mamyshev oscillator}
\subsection{Method}
Pulse propagation in the Mamyshev oscillator cavity is modeled using the generalized nonlinear Schr\"odinger equation (GNLSE):
\begin{equation}
\label{eq:GNLSE}
\frac{\partial A}{\partial z} = -\frac{i\beta_2}{2}\frac{\partial^2 A}{\partial t^2} + \frac{g-\alpha}{2} A + i\gamma|A|^2 A,
\end{equation}
where $A(z,t)$ is the complex envelope of the intracavity optical field, $\beta_2$ is the group velocity dispersion for the waveguide, $\gamma$ is the waveguide Kerr nonlinear coefficient detailed later, $g(z)$ is the position-dependent optical gain, and $\alpha$ is the linear optical loss. 
Propagation is performed sequentially in the forward and backward directions using the split-step Fourier method~\cite{agrawalnonlinear}, assuming negligible nonlinear interaction between counter-propagating pulses.
Accurate modeling of $g(z)$ requires additional effort and we adopt a first-principles static model instead of the gain saturation model commonly used in literature. 
The latter assumes a fixed small-signal gain $g_0$ and a constant saturation power $P_\mathrm{sat}$, with the gain given by $g = \frac{g_0}{1 + P/P_\mathrm{sat}}$.
In erbium-doped waveguides with sub-meter-scale lengths and high doping levels, pump power can decay significantly along the propagation direction, which makes the assumption of a constant $P_\mathrm{sat}$ inaccurate.\footnote{\fntxt}
To address this, we aim to model the non-constant population inversion $n(z)$ and gain $g(z)$ for each point during the pulse and pump propagation, using the effective two-level system rate equations \cite{BECKER1999xv, giles2002modeling}:

\begin{equation}
\label{eq:rate}
\begin{aligned}
\frac{\dd N_2}{\dd t} &= - \frac{N_2}{\tau} 
+ \left( N_1 \sigma_{12\mathrm{s}} - N_2 \sigma_{21\mathrm{s}} \right) \phi_{\mathrm{s}} 
- \left( N_2 \sigma_{21\mathrm{p}} - N_1 \sigma_{12\mathrm{p}} \right) \phi_{\mathrm{p}} \\
\frac{\dd N_1}{\dd t} &= \frac{N_2}{\tau} 
+ \left( N_2 \sigma_{21\mathrm{s}} - N_1 \sigma_{12\mathrm{s}} \right) \phi_{\mathrm{s}} 
- \left( N_1 \sigma_{12\mathrm{p}} - N_2 \sigma_{21\mathrm{p}} \right) \phi_{\mathrm{p}}.
\end{aligned}
\end{equation}
Under the assumption that the population inversion has reached equilibrium ($\dd N_{1,2}/\dd t=0$) and the inversion is uniform across the waveguide cross-section, the normalized population inversion $n(z)=N_2/N'_0$ becomes
\begin{equation}
\label{eq:N2}
n=N_2/N_0' = 
\frac{
\tau \frac{\sigma_{12\mathrm{s}}}{h \nu_{\mathrm{s}}} I_{\mathrm{s}} + \tau \frac{\sigma_{12\mathrm{p}}}{h \nu_{\mathrm{p}}} I_{\mathrm{p}}}{
\tau \frac{\sigma_{12\mathrm{s}} + \sigma_{21\mathrm{s}}}{h \nu_{\mathrm{s}}} I_{\mathrm{s}} + \tau \frac{\sigma_{12\mathrm{p}} + \sigma_{21\mathrm{p}}}{h \nu_{\mathrm{p}}} I_{\mathrm{p}} + 1
}.
\end{equation}

The resulting gain coefficient $g(z)$ and pump power evolution are:
\begin{equation}
\label{eq:gain}
\begin{aligned}
g(z) &= (n(z) (1-n_\mathrm{cluster})\sigma_{21\mathrm{s}} - ((1-n(z))(1-n_\mathrm{cluster})+n_\mathrm{cluster}) \sigma_{12\mathrm{s}})N_0\Gamma\frac{n_g}{n_A} \\
\frac{\dd I_{\mathrm{p,prop}}(z)}{\dd z} &= \left( n(z) (1-n_\mathrm{cluster})\sigma_{21\mathrm{p}} - ((1-n(z))(1-n_\mathrm{cluster})+n_\mathrm{cluster}) \sigma_{12\mathrm{p}} \right)N_0\Gamma\frac{n_g}{n_A} I_{\mathrm{p,prop}}(z) - \alpha I_{\mathrm{p,prop}}(z),
\end{aligned}
\end{equation}
where the definition of parameters in Eqs. \eqref{eq:rate}-\eqref{eq:gain} can be found below: 
\begin{itemize}[itemsep=0.5pt,parsep=0pt]
    \item $\sigma_\mathrm{12s}$ and $\sigma_\mathrm{12p}$: absorption cross sections for signal and pump.
    \item $\sigma_\mathrm{21s}$ and $\sigma_\mathrm{21p}$: emission cross sections for signal and pump.
    \item $\phi_\mathrm{s}$, $\phi_\mathrm{p}$: photon fluxes for signal and pump (sum of both forward and backward propagating light).
    \item $I_\mathrm{s}$ and $I_\mathrm{p}$: optical intensities of signal and pump (sum of both forward and backward propagating light).
    \item $\alpha$: power loss coefficient in either forward or backward propagation, assumed to be the same for pump and signal.
    \item $g$: power gain coefficient for signal in either forward or backward propagation.
    \item $I_\mathrm{p,prop}$: optical intensity of the pump in either forward ($I_\mathrm{p,fwd}$) or backward ($I_\mathrm{p,bwd}$) propagation. 
    \item $\tau$: spontaneous emission lifetime.
    \item $h \nu_{\mathrm{s}}$, $h \nu_{\mathrm{p}}$: photon energies of signal and pump.
    \item $N_0$: peak erbium ion concentration.
    \item $\Gamma$: overlap factor between the erbium ion distribution and optical mode intensity, normalized to the peak concentration.
    \item $N_1(z)$, $N_2(z)$: ground- and excited-state population densities (peak of the profile).
    \item $N_0' = N_1(z) + N_2(z)$: total active erbium concentration (peak of the profile).
    \item $n_\mathrm{cluster} = 1 - N_0' / N_0$: fraction of erbium ions in ion clusters that cannot be excited, while still absorbing.
    \item $n_g/n_A$: correction factor of the gain in high-index-contrast waveguides \cite{Robinson:08}.
\end{itemize}

In the gain simulation, we assumed a uniform excitation ratio in the cross-section of the waveguide, which is a good approximation for high pump and signal intensity.
We considered a simplified phenomenological quenching model similar to the pair-induced quenching in~\cite{Delevaque} due to the relatively high erbium doping concentration (up to 0.5 atomic percent), where some ions are considered to be in "clusters" that absorbs while have negligible radiative emission efficiency.
We did not include uniform up-conversion processes~\cite{Snoeks95CUP}, as these effects were found to be negligible in our strongly pumped system.
This conclusion is based on parameters extracted from separate photoluminescence decay measurements in implanted short waveguide samples.
The emission and absorption cross-sections, as well as the fraction of clustered ions, were empirically determined by fitting the model to measured amplifier gain data.

\begin{algorithm}[H]
\caption{Steady state Mamyshev oscillator simulation}\label{alg:mo}
\KwIn{Guess of the intracavity field: $A(\nu)$, pump power at the boundaries: $P_\mathrm{0, fwd}$, $P_\mathrm{0, bwd}$, complex reflectivity of the two filters: $R_1(\nu)$ and $R_2(\nu)$, and other waveguide properties used in propagation.}

Create initial guesses as function of spatial coordinate $z$ for average power of pulse propagating backwards $P_\mathrm{avg, bwd, 1}(z)$, and pump power propagating backwards $P_\mathrm{p, bwd, 1}(z)$ with piecewise linear functions\;

\For{$j \leftarrow 1$ \KwTo $M$}{
    \tcp{Pseudo-"time stepping" for convergence of spectrum}
    \For{$k \leftarrow 1$ \KwTo $N$}{
        \tcp{Self-consistent iteration for all the optical powers}
        Reset $A(\nu, 0)$ to the $A'(\nu, 0)$ obtained from last "time-stepping"\;
        \For{$m \leftarrow 1$ \KwTo $l/dz$}{
            \tcp{Forward propagation.}
            Current position $z \leftarrow m\times dz$\;
            Compute average signal power at the past position $P_\mathrm{avg, fwd, k}(z-dz)$ using the electric field $A(\nu, z-dz)$\;
            Compute population inversion $n(z)$ using $P_\mathrm{avg, fwd, k}(z-dz)+P_\mathrm{avg, bwd, k-1}(z)$ and $P_\mathrm{p, fwd, k}(z-dz)+P_\mathrm{p, bwd, k-1}(z)$ with Eq. \eqref{eq:N2}\;
            Compute the gain $g(z)$ using $n(z)$ with Eq. \eqref{eq:gain}\;
            Compute the signal electric field $A(\nu, z)$ from $A(\nu, z-dz)$ and $g(z)$ using split-step Fourier method\;
            Compute the pump power $P_\mathrm{p, fwd, k}(z)$ from $P_\mathrm{p, fwd, k}(z-dz)$ using $n(z)$ and propagation equation Eq. \eqref{eq:gain}\;
        }
        \tcp{Apply filter 1.}
        $A(\nu, l)\leftarrow A(\nu,l)\times R_2(\nu)$\; 
        \For{$m \leftarrow 1$ \KwTo $l/dz$}{
            \tcp{Backward propagation.}
            Current position $z \leftarrow l-m\times dz$\;
            Compute average signal power at the past position $P_\mathrm{avg, bwd, k}(z+dz)$ using the electric field $A(\nu, z+dz)$\;
            Compute population inversion $n(z)$ using $P_\mathrm{avg, bwd, k}(z+dz)+P_\mathrm{avg, fwd, k}(z+dz)$ and $P_\mathrm{p, fwd, k}(z+dz)+P_\mathrm{p, bwd, k}(z+dz)$ with Eq. \eqref{eq:N2}\;
            Compute the gain $g(z)$ using $n(z)$ with Eq. \eqref{eq:gain}\;
            Compute the signal electric field $A(\nu, z)$ from $A(\nu, z+dz)$ and $g(z)$ using split-step Fourier method\;
            Compute the pump power $P_\mathrm{p, bwd, k}(z)$ from $P_\mathrm{p, fwd, k}(z-dz)$ using $n(z)$ and propagation equation Eq. \eqref{eq:gain}\;
        }
        \If{$|P_\mathrm{avg, bwd, k}(0)-P_\mathrm{avg, bwd, k-1}(0)|<tol$ or $k=N$}{
            \tcp{Self-consistent iterations converged or exhausted.}            \tcp{Apply filter 2 and prepare for the next "time-stepping" from $z=0$.}
            $A'(\nu, 0)\leftarrow A(\nu, 0)\times R_2(\nu)$\;
            Break\;
        }
    }
}
Postprocess the computed intra-cavity field $A'(\nu, 0)$ to obtain output field and performance estimations\;
\end{algorithm}

Due to the linear cavity design and bidirectional pumping, both signal and pump propagate in forward and backward directions.
Their intensities modify the local population inversion, which in turn affects the gain and absorption experienced during pulse and pump propagation.
Thus, forward and backward propagation and the propagation of pulse and pump are interdependent and must be solved in a self-consistent manner.

In this work, we searched for steady-state solutions of the system using Algorithm \ref{alg:mo}, where the average powers are assumed to be time-independent and the average powers obtained from pulse propagation (accounting for gain and loss) are consistent with those used in the gain calculation.
This avoids simulating the slow population dynamics (microsecond timescale) over many cavity roundtrips (nanosecond timescale), which would otherwise be computationally expensive.
A ``seed'' pulse is used as the initial guess of the intracavity field, but it does not necessarily represent the actual seed pulse used to initiate the mode-locking.
This ``seed'' is propagated in both forward and backward directions of the cavity consequently.
For each propagation step at position $z$, we compute the steady-state population inversion $n = N_2 / N_0'$ using Eq. \eqref{eq:N2} and the intensities $I_\mathrm{s}$ and $I_\mathrm{p}$ as detailed in the Algorithm \ref{alg:mo}.
Then we compute $g(z)$ using Eq. \eqref{eq:gain} and proceed with the split-step stepping and the scalar propagation of the pump.
The average signal and pump powers in both directions are stored as guesses for the following reversed propagation step.
This bidirectional propagation loop is repeated iteratively until the forward and backward power profiles converge.
Once convergence is achieved, the updated intracavity field $A$ is used as the input for the next outer iteration (“pseudo-time step”) to refine the pulse shape and spectrum.
The overall process continues until a given number of outer loops is reached.

The algorithm typically converges within tens of outer iterations across most of the parameter space, either settling into a stable mode-locked state or collapsing to a trivial solution dominated by continuous-wave lasing from parasitic reflections.
In some cases, the algorithm fails to converge, which we attribute to chaotic pulse dynamics which have also been occasionally observed experimentally as unstable or chaotic pulsing behavior under specific pump powers and grating separations.

\subsection{Parameters}
\begin{figure}[htbp]
\centering
\includegraphics[width=0.7\linewidth]{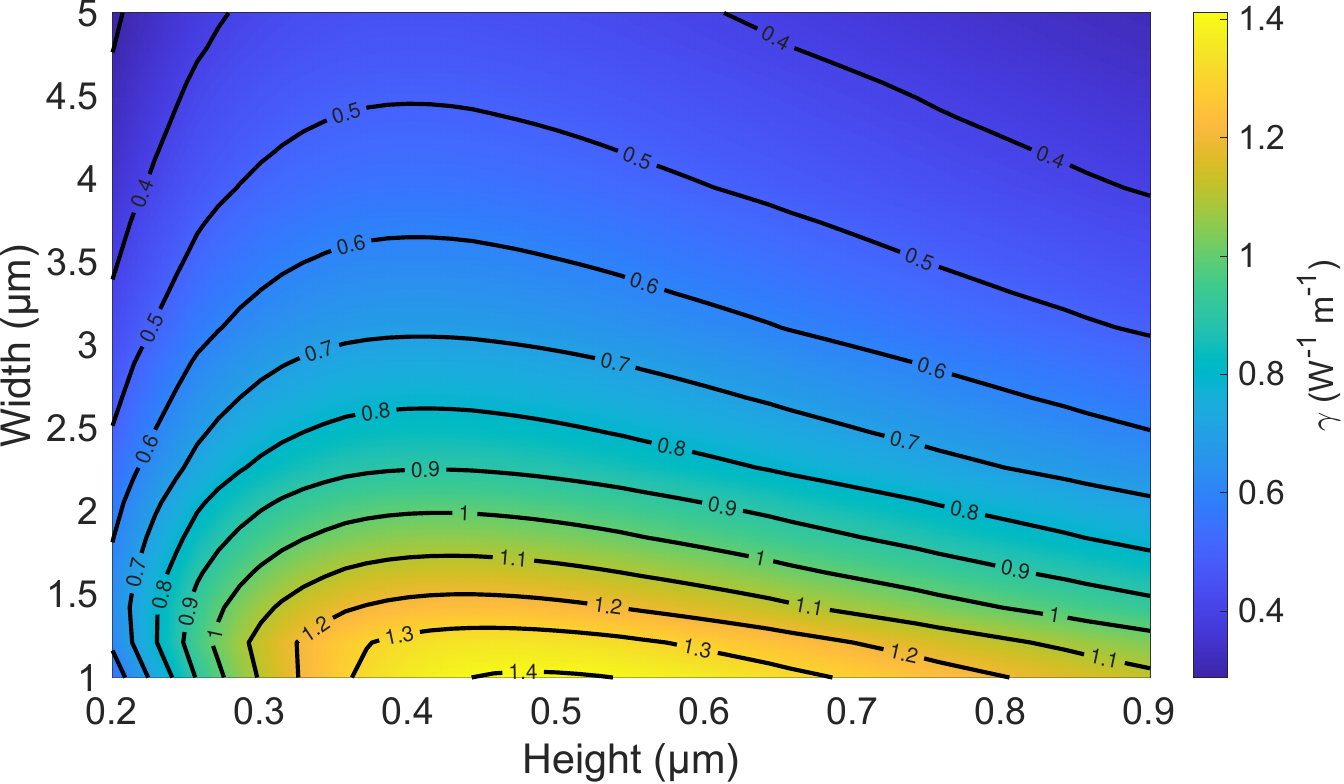}
\caption{Computed Kerr nonlinear coefficient $\gamma$ as a function of the width and height of a \ce{Si3N4} waveguide at \SI{1.55}{\micro\meter} wavelength.}
\label{fig:gamma}
\end{figure}

To estimate the Kerr nonlinear coefficient $\gamma$ of our \ce{Si3N4} waveguide, we use the definition:
\begin{equation}
\gamma = \frac{2\pi}{\lambda}\frac{n_2}{A_{\mathrm{eff}}},
\end{equation}
where $\lambda= 1.55~{\mathrm{\mu m}}$ and $n_2 = 2.2\times 10^{-19}\,\mathrm{m^2/W}$ is the Kerr nonlinear index of \ce{Si3N4}~\cite{gao2022probing}. 
The nonlinear contribution from the \ce{SiO2} cladding is neglected due to the significantly lower $n_2$ in \ce{SiO2} and the confinement of the optical mode. 
$A_{\mathrm{eff}}$ is the effective mode area of the waveguide, which can be calculated based on a fully vectorial model~\cite{ShahraamAfshar:13}:
\begin{equation}
\quad A_{\mathrm{eff}} = \frac{\mu_0}{\varepsilon_0} \frac{3\left|\int_{\infty} \left(\mathbf{E} \times \mathbf{H}^*\right) \cdot \dd\boldsymbol{S} \right|^2}{n_{\mathrm{Si_3N_4}}^2 \int_{\mathrm{Si_3N_4}} \left[ 2|\mathbf{E}|^4 + \left| \mathbf{E}^2 \right|^2 \right] \dd S},
\end{equation}
where $\mathbf{E}$ and $\mathbf{H}$ are the modal electric and magnetic fields derived from finite element simulation with COMSOL, $n_\mathrm{Si_3N_4}$ is the refractive index of \ce{Si3N4} and $\dd\boldsymbol{S}$ is the vectorial surface element of the cross-section.
Figure~\ref{fig:gamma} shows the calculated values of the nonlinear coefficient $\gamma$ for the fundamental transverse electric (TE) mode across different waveguide geometries, as this mode typically exhibits the strongest confinement and highest nonlinearity.
In this work, we selected a \ce{Si3N4} waveguide thickness of 350~nm as a compromise between achieving high nonlinearity and ensuring compatibility with cost-effective ion implantation using air-insulated electrostatic accelerators.

The complex reflectivity $R_1(\nu)$ and $R_2(\nu)$, as well as the transmissivity of the waveguide Bragg gratings (WBGs), were simulated using the transfer matrix method (TMM) \cite{bjork1987new}.
The effective index modulation $\Delta n$ and parasitic chirping $\Delta l/l$ caused by the bandgap center shift due to apodization~\cite{bruckerhoff2025general}, were extracted from experimentally measured grating spectra fabricated during calibration runs. 
We note that the experimentally extracted $\Delta n$ in our gratings can be lower than the numerical predictions, by either the effective index contrast or photonic bandgap simulations~\cite{bruckerhoff2025general}, even after compensating for the finite lithographic resolution.
This discrepancy likely arises from the breakdown of the perturbative approximation for the strongly modulated gratings used here.
Nevertheless, we find that TMM provides a good approximation of the grating spectrum after using the experimentally calibrated $\Delta n$.

Mamyshev oscillator MLLs are particularly sensitive to parasitic backreflections, which can lead to undesirable continuous-wave lasing in the cavity, reduced round-trip gain, and eventually failure to reach stable mode-locking.
To model wideband Fresnel backreflections at the chip facets, which represent the dominant source of reflections, we introduce a correction term to the grating reflectivity: $R_i'=R_i+\sqrt{R_\mathrm{parasitic}}(1-|R_i|^2)\exp(\mathrm{i}\varphi)/[1+R_i\sqrt{R_\mathrm{parasitic}}\exp(\mathrm{i}\varphi)]$, where $R_\mathrm{parasitic}$ is the measured power backreflection ratio, and $\varphi$ is the frequency-dependent, round-trip propagation phase between the grating and facet.

The typical parameters for the integrated Mamyshev MLL is listed in Table \ref{tab:physparams}.
These parameters were used to produce the pulse propagation simulation shown in Fig. 1(c) of the main text.
The silicon nitride waveguide is assumed to have an refractive index of 1.98 at the interested wavelengths.

\begin{longtable}{@{}llll@{}}
\caption{Parameters used in the integrated Mamyshev MLL simulation}
\label{tab:physparams} \\
\toprule
{Symbol} &{Variable name} & {Description} & {Value}\\
\midrule
\endfirsthead
\multicolumn{4}{l}{\small\itshape Table~\thetable\ (continued)}\\
\toprule
{Symbol} &{Variable name} & {Description} & {Value}\\
\midrule
\endhead
\bottomrule
\endfoot
$\lambda_s$ &  & Simulation center wavelength & $1550\,\mathrm{nm}$\\
$\lambda_p$ &  & Pump wavelength              & $1480\,\mathrm{nm}$\\
$l$ & \texttt{l}            & Length of doped waveguide & $0.42\,\mathrm{m}$\\
$n_g$ & \texttt{ng}           & Waveguide group refractive index & 2.0023\\
$\beta_2$ & \texttt{beta2}        & Group-velocity dispersion & $0.715\,\mathrm{ps^{2}/m}$\\
$\beta_3$ & \texttt{beta3}        & Third-order dispersion & $-1.81\times10^{-3}\,\mathrm{ps^{3}/m}$\\
$\gamma$ & \texttt{gamma\_nl}    & Kerr nonlinear coefficient & $1.145\,\mathrm{W^{-1}\,m^{-1}}$\\
$\alpha$ & \texttt{alpha\_loss}  & Propagation loss (from section \ref{ss:loss})& $10\,\mathrm{dB/m}=2.3\,\mathrm{m^{-1}}$\\
$n_\mathrm{cluster}$ & \texttt{clusteringFrac} & Fraction of clustered ions & 0.06\\
$\Gamma$ & \texttt{overlap}        & Mode-gain overlap & 0.22\\
$N_0$ & \texttt{N0}             & Peak total erbium ion concentration & $3.68\times10^{26}\,\mathrm{m^{-3}}$\\
$A_s$, $A_p$ & \texttt{As}, \texttt{Ap} & Effective mode area (signal, pump) & $1.5\times10^{-12}\,\mathrm{m^{2}}$\\
$\tau$ & \texttt{tau}            & Upper-state lifetime & $3.0\times10^{-3}\,\mathrm{s}$\\
$\sigma_\mathrm{12p}$, $\sigma_\mathrm{12s}$ & \texttt{s12p}, \texttt{s12s} & Absorption cross-sections (pump, signal) & $3.43\times10^{-25}\,\mathrm{m^{2}}$ ; $2.53\times10^{-25}\,\mathrm{m^{2}}$\\
$\sigma_\mathrm{21p}$, $\sigma_\mathrm{21s}$ & \texttt{s21p}, \texttt{s21s} & Emission cross-sections (pump, signal) & $5.9\times10^{-26}\,\mathrm{m^{2}}$ ; $4.48\times10^{-25}\,\mathrm{m^{2}}$\\
$\Delta \lambda_g$ & \texttt{delta\_lambda\_g} & Gain bandwidth & $40\,\mathrm{nm}$\\
$P_\mathrm{p, fwd}$ & \texttt{pump\_power\_fwd} & On-chip forward pump power & $450\,\mathrm{mW}$\\
$P_\mathrm{p, bwd}$ & \texttt{pump\_power\_bwd} & On-chip backward pump power & $390\,\mathrm{mW}$\\
$\Delta\lambda_\mathrm{f}$ & \texttt{filter\_gap}   & Offset between filters & $8.9\,\mathrm{nm}$\\
$R_\mathrm{parasitic}$ & \texttt{br}           &Power back reflection ratio & $1.0\times10^{-3}$\\
$N$ & \texttt{NN}   & Number of grating periods & $950$\\
$\Delta n$ & \texttt{dn}           & Peak to peak effective index modulation & $8.6\times10^{-3}$\\
$\Delta l/l$ & \texttt{parasitic\_chirping}  & Parasitic chirping due to index change & $1.2\times10^{-3}$\\
$\sigma_\mathrm{apo}$ & \texttt{apodization\_sigma}  & Variation for grating's Gaussian apodization & $0.15$\\
\end{longtable}

\subsection{Results}
\begin{figure}[htbp]
\centering\includegraphics[width=0.9\linewidth]{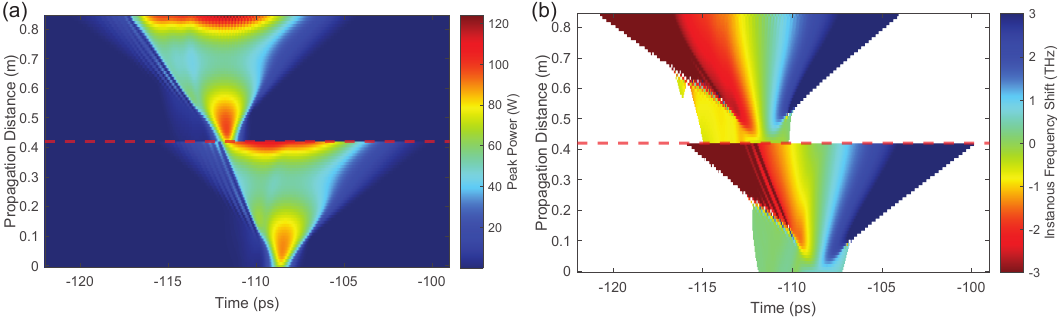}
    \caption{\textbf{Pulse propagation in the time domain.}
    (a) Time-domain intensity of the pulse as a function of propagation length within the cavity.
    The red dashed line indicates the transition from forward to backward propagation after filtering.
    (b) Instantaneous frequency of the pulse as a function of propagation distance.
    The white area indicates where the pulse intensity is less than $0.1\%$ of the peak and the chirping is irrelevant.}
    \label{figS:tdprop}
\end{figure}

For reference, we define Grating 1 as the red-shifted WBG and Grating 2 as the blue-shifted one.
``Forward propagation'' refers to the direction from Grating 1 to Grating 2, while ``backward propagation'' is from Grating 2 to Grating 1.
All reported powers in the simulation correspond to on-chip values.

Figure~\ref{figS:tdprop} shows the simulated time-domain evolution of the pulse as it propagates through the cavity.
The pulse acquires a linear chirp and broadens significantly to several picoseconds due to the strong normal group velocity dispersion (GVD) of the waveguide, which disperses the newly generated frequency components.
The intracavity peak power remains within 50~W to 125~W throughout the round-trip, allowing for substantial self-phase modulation (SPM) without inducing pulse instability or damage to the waveguide.

\begin{figure}[htbp]
\centering\includegraphics[width=1\linewidth]{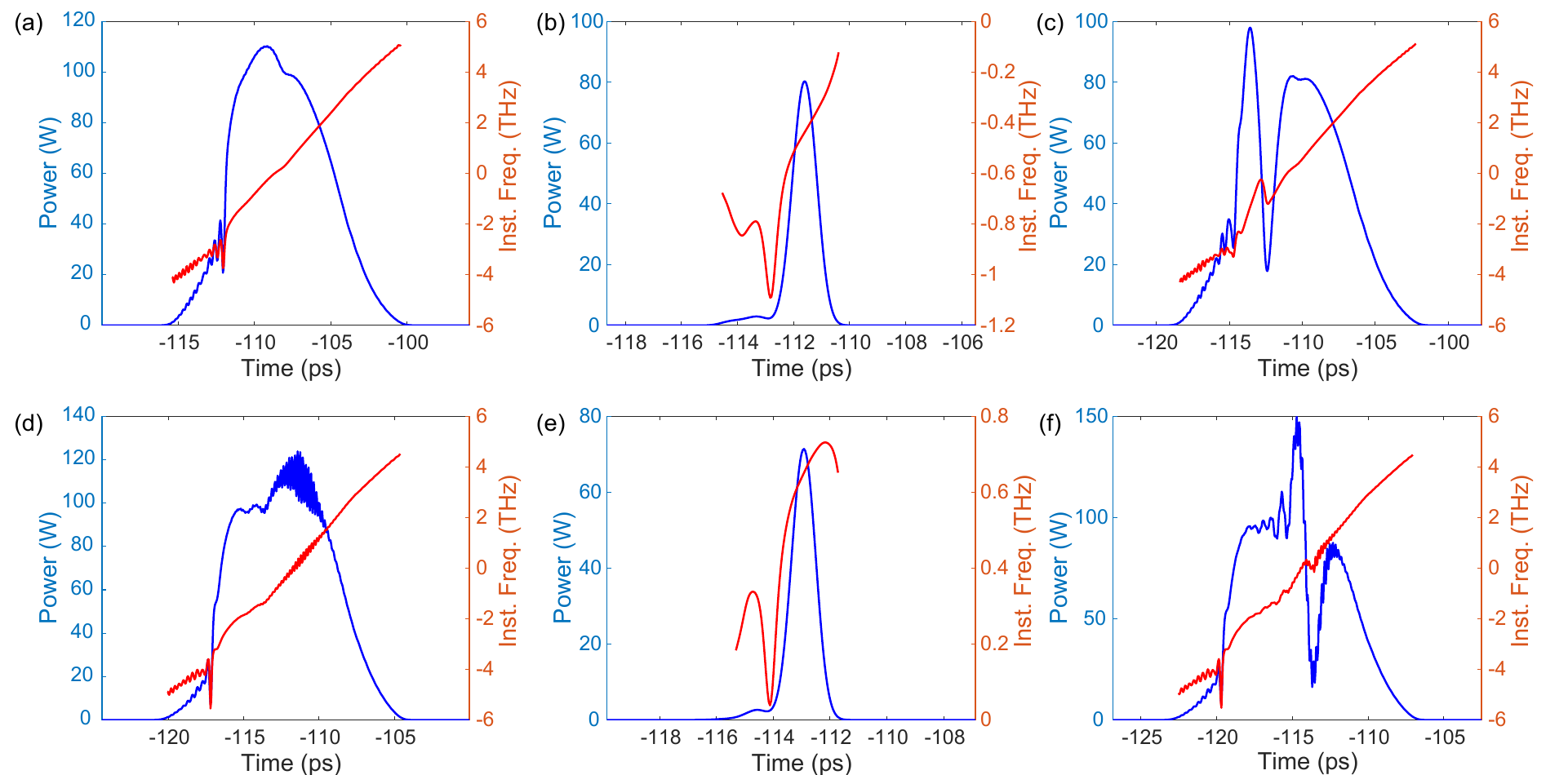}
    \caption{\textbf{Pulse waveforms and chirp at key locations.}
    (a) Pulse after forward propagation, before WBG reflection.
    (b) Pulse after forward propagation, immediately after reflection.
    (c) Output pulse transmitted through WBG after forward propagation.
    (d-f) Same as (a-c), but for backward propagation.
    }
    \label{figS:tdout}
\end{figure}

\begin{figure}[h]
\centering\includegraphics[width=0.8\linewidth]{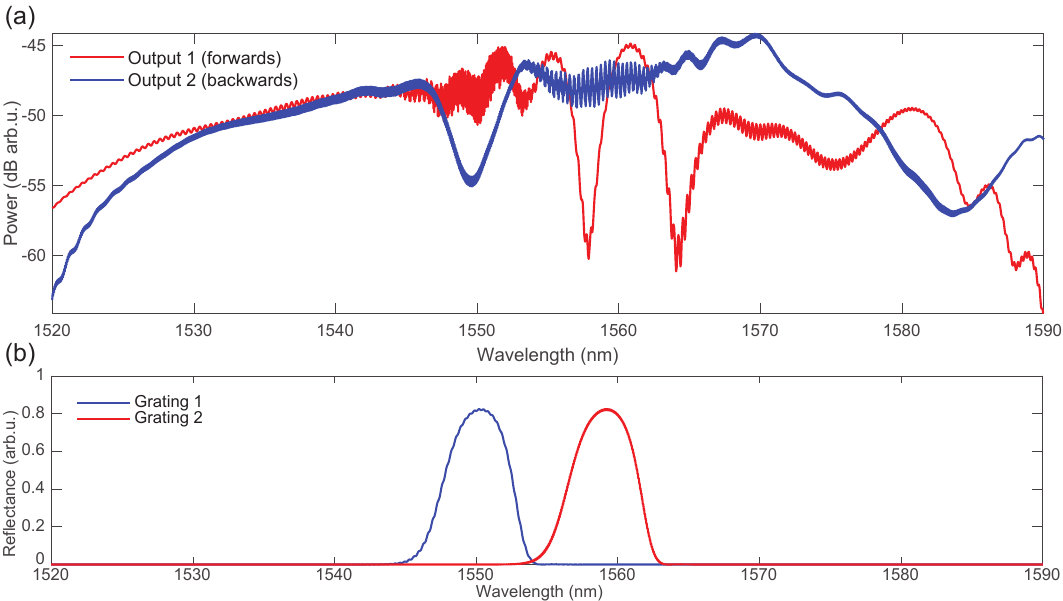}
    \caption{\textbf{Simulated frequency-domain results.}
    (a) Output spectra of the Mamyshev oscillator.
    (b) Reflectance spectra of the WBGs including modeled facet parasitic reflections.}
    \label{figS:fdout}
\end{figure}

Figure~\ref{figS:tdout} shows the time-domain envelope and instantaneous frequency at various points in the cavity, including the reflected and transmitted pulses at the WBGs.
The simulated output pulses from both gratings feature pulse widths of a few picoseconds and clearly exhibit a linear frequency chirp.
This indicates that the output pulses can be significantly compressed by group delay dispersion compensation. 
Figure~\ref{figS:fdout} presents the simulated optical spectra of the laser outputs along with the modeled WBG reflectance.
While the details of the spectra differed from the measured ones, the simulations reproduced key features such as spectral valleys and fine oscillations on one side of the output.

\begin{figure}[htbp]
\centering\includegraphics[width=0.9\linewidth]{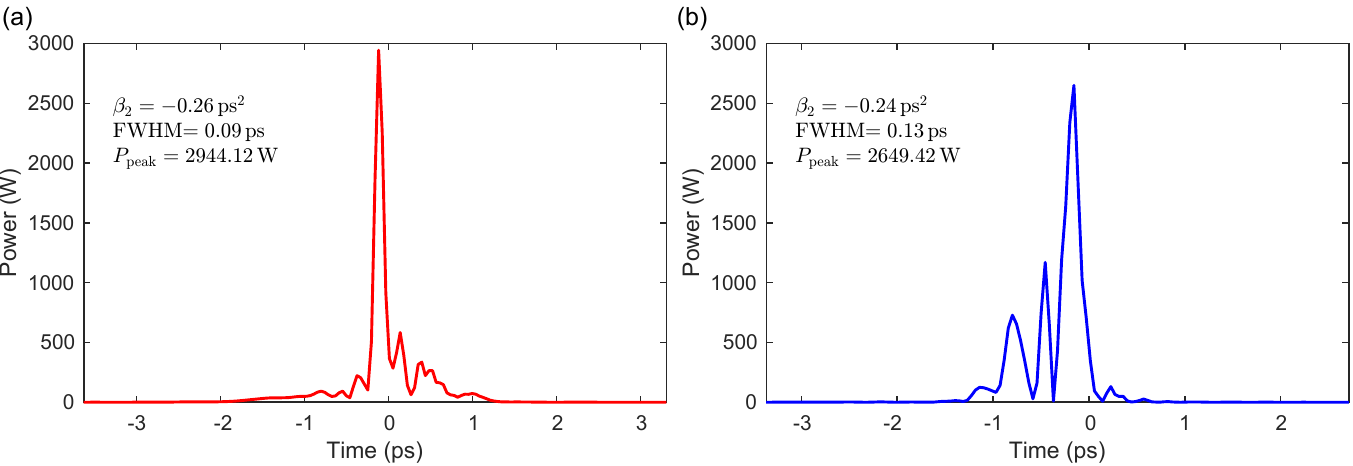}
    \caption{\textbf{Compressed output pulses.}
    (a) Forward-propagated and (b) Backward-propagated output pulse after group delay dispersion compensation.
    $\beta_2$: group delay dispersion applied.
    FWHM: full width at half maximum of the compressed pulse.
    $P_\mathrm{peak}$: peak power after compression.}
    \label{figS:tdcompress}
\end{figure}

Figure~\ref{figS:tdcompress} presents the pulse after optimal group delay dispersion compensation.
Sub-150~fs compressed pulses are achieved from both output directions, with peak powers in the kilowatt levels.

\begin{figure}[htbp]
\centering\includegraphics[width=0.8\linewidth]{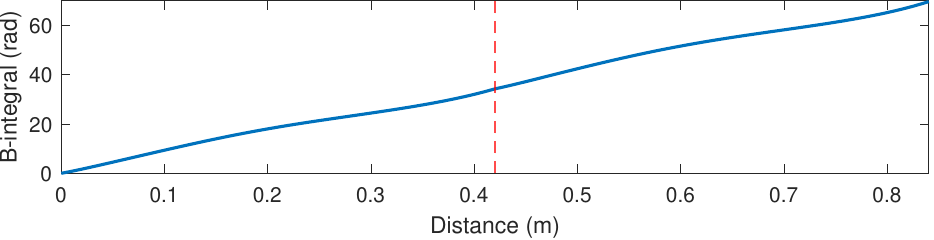}
    \caption{Accumulated nonlinear phase shift (B-integral) during pulse propagation in the linear cavity.
    The vertical line separates forward and backward propagation segments.}
    \label{figS:Bint}
\end{figure}

The accumulated nonlinear phase shift during a round-trip is semi-quantitatively characterized using the B-integral $B = \int_0^{2l} {\gamma I_\mathrm{s,prop}(z)\,\dd z}$~\cite{fu2018several}.
Figure~\ref{figS:Bint} shows the B-integral as a function of propagation length. The maximum B-integral reaches approximately $20\pi$ per round-trip, which is well beyond the regime accessible in soliton-based mode-locking~\cite{liu2017megawatt, fu2018several}.

\begin{figure*}[htbp]
\centering
\includegraphics[width=1\linewidth]{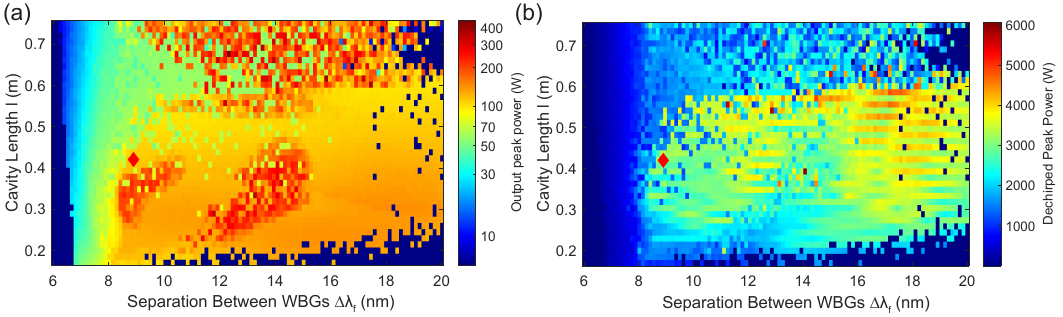}
\caption{\textbf{Parameter map of peak power.}
(a) Intracavity peak power after backward propagation as a function of WBG center wavelength offset $\Delta\lambda_\mathrm{f}$ and cavity length $l$.
Color map is in logarithmic scale.
(b) Output peak power after optimal group delay dispersion compensation.
The red diamond marker indicates the location of the current design.}
\label{figS:mappp}
\end{figure*}

Figure~\ref{figS:mappp} shows the simulated intracavity and dispersion compensated output peak power of the stable solution as a function of two key design parameters: the grating center wavelength offset $\Delta\lambda_\mathrm{f}$ and the cavity length $l$.
A broad parameter space supports mode-locking, with WBG separations ranging from 7~nm to 20~nm and cavity lengths from 20~cm to 70~cm.
However, for cavity lengths exceeding 57~cm combined with relatively large grating separations, the algorithm often fails to converge to a stable solution.
These regions are suspected to correspond to chaotic states where the pulse becomes unstable, possibly due to excessive nonlinear phase accumulation within a round trip, leading to instability.
Our current design operates with a moderate cavity length and relatively small grating separation, well within the stable regime and have margin for fabrication errors and future development for different repetition rates.

\FloatBarrier

\section{Detailed description of the experiment setup}
\subsection{Basic mode-locking operation}

We used two high-power, single-mode silica fiber (SMF)-coupled 1480nm pump lasers in the experiment: a Connet VLSS-1480-B-800-1 (Pump 1 in the main text) and a Lumibird CRFL-05-1480-OM1-8230-FA (Pump 2).
The Connet laser employs a proprietary architecture, while the Lumibird laser is a Raman fiber laser.
When the 959~mW pump 2 is used, the Lumibird laser was operated at 2~W output power and attenuated before sending to the wavelength division multiplexer (WDM).
For future integration, commercially available high power indium phosphide laser diodes~\cite{lumentum1480diodes} can be used instead.
Two thin-film-filter-based WDMs (Ascentta FWDM-45-L-10-FA) were used to inject the 1480~nm pump light while simultaneously extracting the laser output in the 1550~nm band.
To minimize back-reflection from the pump lasers, we selected filter-based WDMs with a ``1480nm pass / 1550nm reflect'' configuration, which offer better 1550~nm band isolation at the 1480~nm port compared to typical fused-fiber WDMs or the ``1550nm pass / 1480nm reflect'' variants.
These WDMs also feature sufficiently wide passbands in the 1550~nm range to accommodate the MLL output spectrum.
The pump power after the WDM was calibrated using an optical power sensor (Thorlabs S145C).
Isolators or circulators were placed along the 1550~nm signal path to suppress back-reflections from photodetectors and other instruments.
The integrated microheaters were powered by a DC power supply (Rigol DP832A).
For many samples, stable mode-locking could be reached without using the microheaters although some required tuning to prevent parasitic continuous-wave lasing.
Light was coupled into and out of the chip by butt-coupling with cleaved Coherent UHNA7 fibers.
To reduce facet reflections, the fiber-chip interface was covered with index-matching gel (Thorlabs G608N3), ensuring no air gap remained at the contact point that could otherwise lead to excessive back-reflection.
The sample was mounted on a metal block equipped with a thermoelectric cooler (TEC), which was feedback-controlled via a thermistor for temperature stabilization.
Thorlabs NanoMax flexure stages were used to position the fibers.
Average output powers from both outputs were monitored using Thorlabs~S144C power sensors, placed after several broadband fiber directional couplers.
During calibration, the wavelength-dependent insertion loss between the common port of the WDMs and the power sensors was characterized using a swept tunable laser and taken into account to ensure accurate measurement of the laser output power.
The optical spectra from both outputs were simultaneously monitored using two optical spectrum analyzers (OSAs): a Yokogawa~AQ6370D with a 0.2~nm resolution (Output 1), and a HP~71451B with 1~nm resolution (Output 2).
The power spectral density (PSD) plots shown in the main text were corrected for insertion loss and normalized to a resolution bandwidth of 0.2~nm for both outputs.
However, there remains significant uncertainty in the absolute power level in the optical spectrum due to the varying physical resolution bandwidths and instrument bandpass of these grating-based OSAs.
The time-domain pulse traces were recorded using a DC-coupled fast photodiode (Thorlabs DET08CFC/M) and a digital oscilloscope (Keysight EXR608A) set to 50~$\Omega$ input impedance.

\subsection{Autocorrelation characterization}

To characterize the pulse duration after applying dispersion compensation, we performed intensity autocorrelation measurements using the output from Port 1 of the laser (Path A in the main text).
The signal was first sent through an optical isolator to prevent backreflections, then passed through a programmable optical filter (Finisar Waveshaper 1000s) for dispersion compensation, and subsequently amplified using a home-built erbium-doped fiber amplifier (EDFA).
The total fiber path length from the chip facet to the autocorrelator (APE pulseCheck USB 50) was approximately 11~m.

The programmable filter introduces an insertion loss of about 6\,dB, which was compensated by the EDFA.
We avoided the use of commercial EDFAs, which typically employ several meters of erbium-doped fiber and are known to introduce distortions to the pulse due to uncontrolled dispersion and nonlinear effects at high peak powers.
Instead, we constructed an EDFA with minimal fiber length to preserve the pulse fidelity.
The custom EDFA consisted of two 980/1550~nm WDMs (Thorlabs), a 980~nm pump laser diode (Aerodiode), and a short ($\sim$50~cm) section of highly doped erbium fiber (LIEKKI ER80-4/125-HD-PM).
The doped fiber used is polarization-maintaining, although this property was not needed and the selection was based on stock availability.
To further minimize unwanted nonlinear broadening, we ensured that the fiber connection between the EDFA output and the autocorrelator was kept as short as possible (approximately 1~m).
The output power of the EDFA was adjusted by tuning the pump diode current, typically set to deliver 30 to 50~mW of average power.

In a different approach, we employed the group delay dispersion of the SMF to compress the MLL output pulse. 
The signal from output 2 of the laser (Path B) was routed through a SMF delay line to the autocorrelator.
In this case, no EDFA was used and the autocorrelation measurement was performed directly after the fiber.
The total length of the fiber is approximately 10~m, which correspond to a dispersion of $-0.22~\mathrm{ps}^2$.

\subsection{Coherence characterization}
To characterize the radio-frequency (RF) repetition-rate beatnote via direct photodetection, we used a high-power-handling PIN photodiode (Discovery Semiconductor DSC40S).
The RF spectrum was measured using an electronic signal analyzer (ESA, Keysight N9020A), while phase noise was characterized using a signal source analyzer (Rohde \& Schwarz FSUP26), which offers sensitivity beyond the phase noise floor of standard ESAs.
The timing jitter $T_\mathrm{j}$ was calculated by integrating the measured phase noise spectrum (after removing the spurs), using $T_\mathrm{j} = \frac{1}{2\pi f_\mathrm{rep}} \sqrt{2 \int 10^{\mathrm{SSBPN}(f)/10} \, \dd f}$, where $\mathrm{SSBPN}(f)$ is the single-sideband phase noise at offset frequency $f$ in dBc/Hz unit.
An optical attenuator (HP 8156A) was used to adjust the optical power incident onto the photodetector.
This adjustment was necessary because the photodetected RF signal phase noise can be affected by photodetector saturation and amplitude-noise-to-phase-noise (AM-to-PM) conversion.
For narrowband measurements of the RF beatnote and its phase noise, we employed bandpass filters (Mini-Circuits ZX75BP-1062-S+ for the 6th harmonic, ZX75BP-188-S+ and SLP-250+ for the fundamental harmonic) and a low-noise RF preamplifier (Mini-Circuits ZFL-1000LN+) to optimize the signal-to-noise ratio.
For wideband spectral analysis of the beatnote, no preamplifier was used to avoid saturation and nonlinear harmonic generation within the amplifier.

To evaluate the linewidth of individual comb lines by heterodyne detection, we first preselected a narrow spectral portion of the MLL output using a tunable fiber Bragg grating (AOS GmbH), aligned to match the wavelength ($\sim1552$~nm) of a single-frequency DFB erbium-doped fiber laser (Koheras Adjustik).
After optimizing polarization, the filtered comb signal and the reference laser were combined using a 50:50 directional coupler and detected with a balanced photodetector (Discovery Semiconductor DSC720-39).
The resulting heterodyne signal was analyzed using an ESA (Rohde \& Schwarz FSW).
For frequency noise extraction, a 100~ms segment of the in-phase (I) and quadrature (Q) baseband signal was recorded from the ESA, converted to phase and post-processed using Welch’s method, with an appropriate Fourier window and time segmentation.

\subsection{Starting by seeding}
\label{ss:seeding}
\begin{figure*}[htbp]
	\centering
	\includegraphics[width=0.9\linewidth]{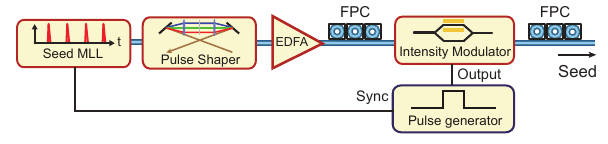}
	\caption{Schematic of the seed source used to initiate Mamyshev oscillator operation.}
	\label{figS:seedsetup}
\end{figure*}

To initiate mode-locking in the integrated Mamyshev oscillator, we used the experimental setup shown in Fig.\,\ref{figS:seedsetup}.
A commercial fiber MLL (Menlo ELMO High Power) served as the seed source, delivering 51~fs full-width-half-maximum (FWHM) pulses at a 100~MHz repetition rate and approximately 100~mW average power. 
The pulse width was considerably shorter than required to start the integrated MLL.
To avoid nonlinear broadening of the seed pulse in subsequent fiber propagation, we employed a programmable optical filter (Finisar Waveshaper 4000S) to reduce the spectrum of the seed pulse using a Gaussian-shaped filter, typically with a bandwidth of 0.15~THz to 0.6~THz.
Meanwhile, a group delay dispersion of $0.4~\mathrm{ps/nm}$ is also programmed to the filter to pre-compensate for dispersion in the fiber-based experiment setup.
and subsequently amplified by a home-built low-nonlinearity erbium-doped fiber amplifier (EDFA), identical to the one used in the autocorrelation measurements.
This compensated for filtering losses and boosted the power to approximately 20~mW.
To select single pulses from the filtered and amplified pulse train, we used an electro-optical Mach-Zehnder intensity modulator (EOSpace) driven by a pulse picker board from Aerodiode.
The pulse picker was synchronized with the seed MLL using the trigger output from the MLL.

The minimum bandwidth of the Gaussian-filtered pulse used for seeding was 0.15~THz, corresponding to a transform-limited pulse width of $\sim$3~ps, which is within reach of integrated III-V mode-locked lasers. 
The seed pulse energy delivered on chip ranged from 15~pJ to 45~pJ, which are also accessible using integrated III-V mode-locked diodes combined with on-chip erbium-doped waveguide amplifiers.
We note that the Mamyshev oscillator is relatively tolerant to variations in seed pulse parameters, particularly when parasitic backreflections from the fiber coupling or experimental setup are minimized.
During experiments, mode-locking can be established across a broad parameter space, even when the seed energy, filter bandwidth, or polarization state were varied substantially. 

\begin{figure*}[htbp]
	\centering
	\includegraphics[width=0.9\linewidth]{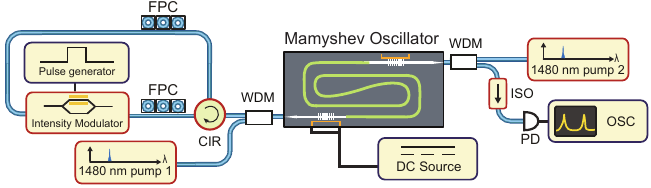}
	\caption{Schematic of the experimental setup for initiating the Mamyshev oscillator using an extended cavity Q-switching method.
	}
	\label{figS:qswsetup}
\end{figure*}
\subsection{Alternative starting method by extended cavity Q-switching}
We observed that a high-energy, nanosecond-duration intracavity pulse can also initiate mode-locking.
To demonstrate this, we constructed an extended cavity by routing one output port of the device back into the waveguide and implemented active Q-switching using an external modulator.
We used a higher off-chip pump power of 2000~mW from Pump 1 and 820~mW from Pump 2 during the starting.
As illustrated in Fig.\,\ref{figS:qswsetup}, a controllable extended cavity was created using an electro-optic intensity modulator (EOSpace) in combination with a circulator (Thorlabs).
The modulator was driven by a function generator (Keysight 33622A) at a repetition rate of 1~kHz, providing high transmission for approximately 200~ns per period in the extended arm of the cavity.
A Q-switch pulse can build up in the main cavity and the extended arm after sending the on pulse, providing a pulse with sufficient energy to initiate the mode-locking (see Section \ref{ss:startingdyn}).
Since the intensity modulator operates on a single polarization with the orthogonal polarization being blocked, we empirically optimized the fiber polarization controller settings by monitoring the Q-switching behavior on an oscilloscope.

\subsection{Supercontinuum generation}
For the supercontinuum generation demonstration, the output from the WDM (including approximately 2.5~m of fiber) was routed through a series of fiber-connected components.
The signal first passed through an optical isolator with a 2.18~m pigtail, followed by a polarization controller (3.43~m of fiber), and was then directed into a 1550~nm lensed fiber (approximately 2~m in length), which coupled the light into a second \ce{Si3N4} chip.
We estimate a coupling loss of 2.5~dB from fiber to chip near 1550~nm.
The optical isolator was included to prevent back-reflection from the lensed fiber and the input facet of the \ce{Si3N4} supercontinuum chip (potentially up to a few percent) which could disrupt stable mode-locking.
For future monolithic implementations, the optical isolator can be omitted, and the approximately 3.8~dB power penalty from coupling loss and fiber components can also be eliminated.
The output from the second chip was collected using a similar 1550~nm lensed fiber (not optimized for short-wavelength coupling efficiency) and routed to two optical spectrum analyzers (Yokogawa AQ6375 and AQ6373) for spectral characterization.
To eliminate potential second-order diffraction artifacts from the grating-based optical spectrum analyzers in the long-wavelength region (\SI{1800}{\nano\meter} to \SI{2400}{\nano\meter}), two Thorlabs FELH1250 long-pass filters were used in a free-space fiber bench (Thorlabs) to block shorter wavelengths during acquisition in this region.

\section{Passive characterization of the device}
\label{ss:loss}
\begin{figure*}[htbp]
    \centering
    \includegraphics[width=0.65\linewidth]{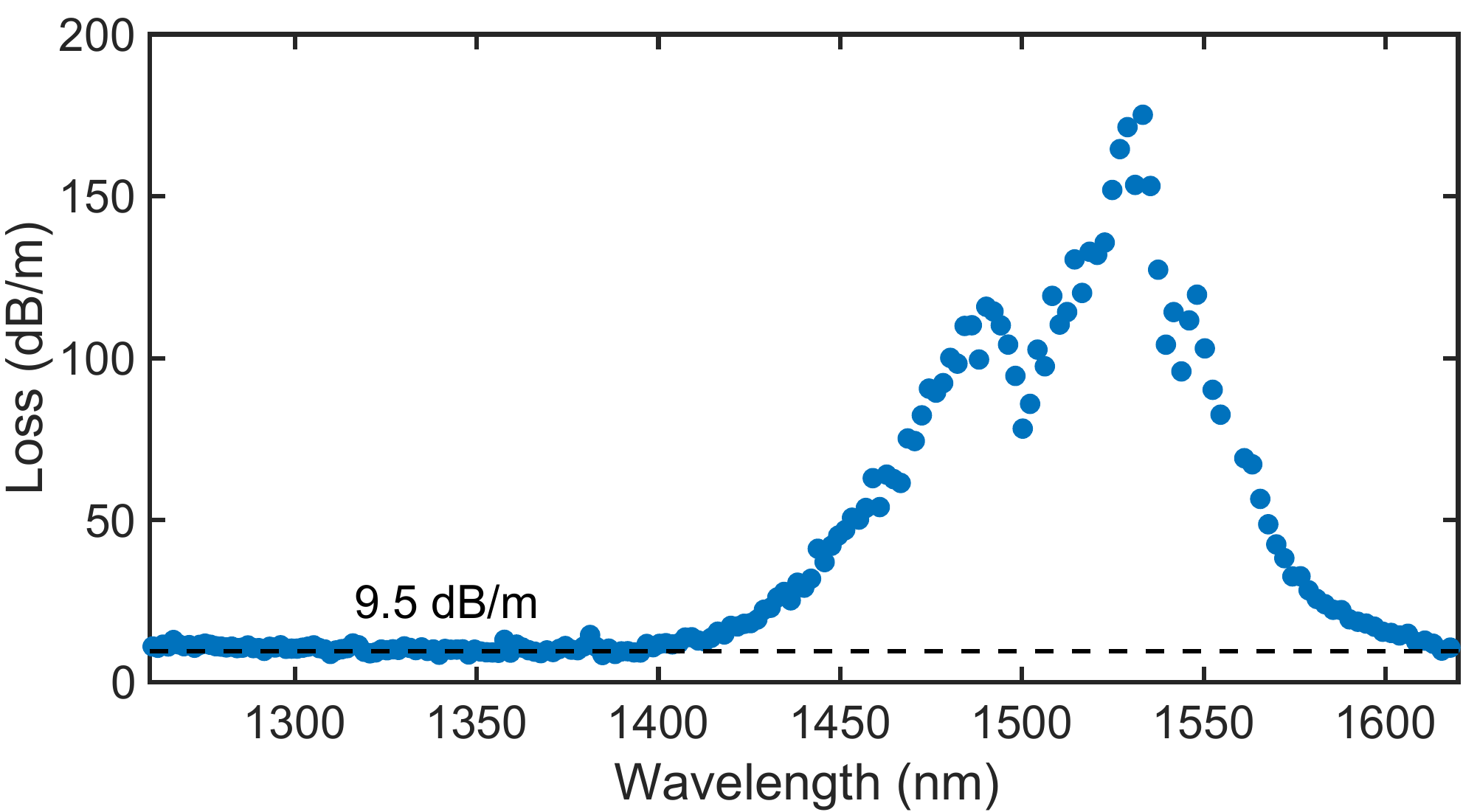}
    \caption{Waveguide loss after ion implantation measured by OFDR.
    The dashed dark line denotes the estimated background loss of approximately 9.5~dB/m, inferred from the measured loss between 1300~nm and 1370~nm.%
    }
    \label{figS:ofdr_loss}
\end{figure*}
We characterized the total waveguide loss of \verb|D20602| samples after ion implantation using optical frequency domain reflectometry (OFDR) with a home-built optical vector network analyzer (described in detail in \cite{riemensberger2022photonic}). 
The waveguide width is around \SI{1.6}{\micro\meter}, and the samples were implanted with a total dose of $3.87\times10^{15}~\mathrm{cm}^{-2}$.
Figure~\ref{figS:ofdr_loss} shows the measured loss spectrum, which reveals a broad erbium absorption spectrum extending from approximately 1400~nm to 1600~nm.
The peak absorbance reaches approximately 165~dB/m at 1533~nm, about twice that of commercial highly erbium-doped fiber (Liekki ER80-8/125).
Outside the erbium absorption band, the background loss of the waveguide is estimated to be approximately 9.5~dB/m.
This relatively low background loss ensures low signal propagation loss and efficient delivery of pump power along the erbium-doped waveguide, leading to high laser efficiency.
Further reduction in loss is expected through improvements in the fabrication process, particularly by eliminating voids on the sides of the waveguide (Supplementary Information \ref{ss:fab}).

The grating reflection spectrum shown in Figure 1(h) of the main text was characterized using a device identical to that used in the main experiments.
A swept single-frequency laser (Toptica CTL) served as the probe source.
The reflected signal was extracted using a 50:50 directional coupler (Fibermart) and measured with a power meter (Thorlabs).
During characterization, the grating microheaters were left unpowered, and the probe polarization was optimized to maximize reflection intensity.
Each of the two gratings was probed individually from the closest ends of the device.
The probe power was kept sufficiently low to avoid reaching the transparency of the erbium-doped waveguides, ensuring that nearly all light was attenuated between the gratings and eliminating interference from the reflection of the opposite grating.
The same UNHA fiber butt-coupling setup as main experiments was used to minimize backreflections from the chip facets during measurement.
\FloatBarrier

\section{Dynamics of pulse build up}
\label{ss:startingdyn}
\begin{figure*}[t]
    \centering
    \includegraphics[width=0.8\linewidth]{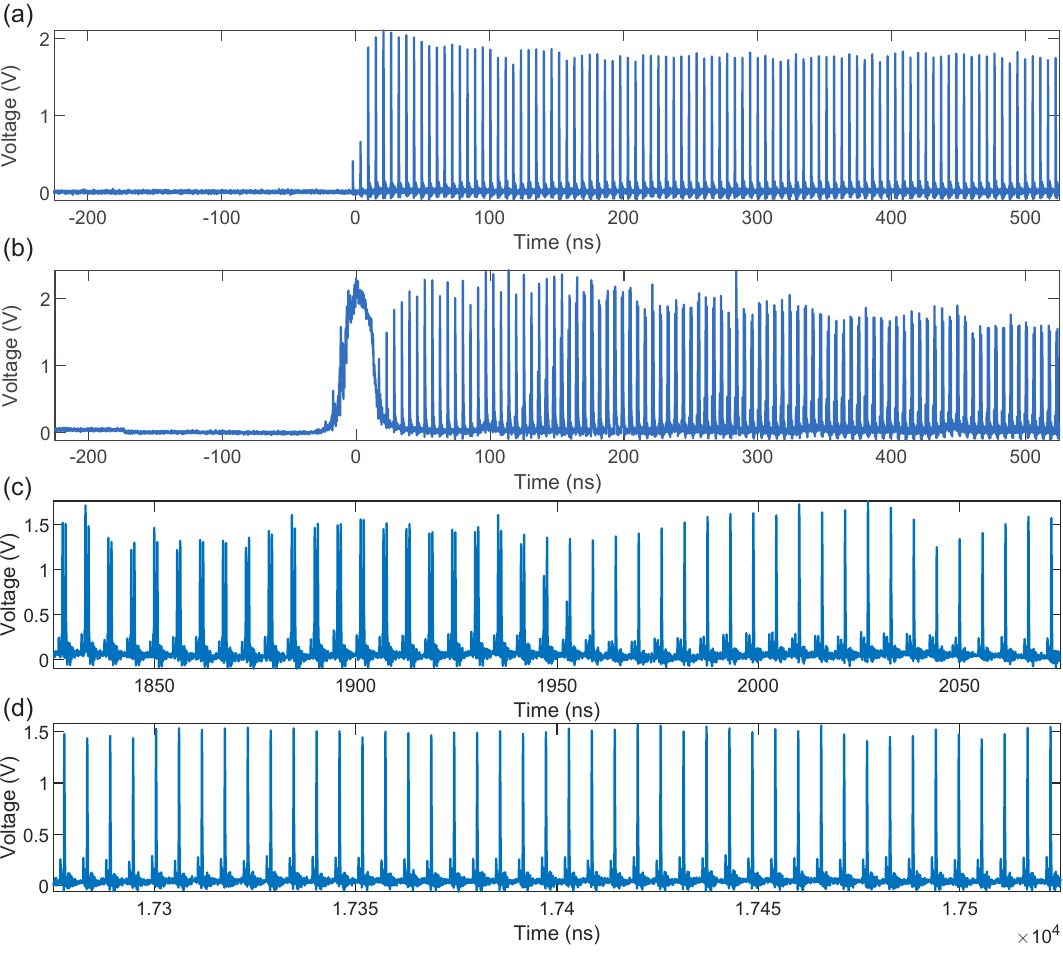}
    \caption{\textbf{Instantaneous output power during the pulse build-up process, monitored with a photodetector and oscilloscope.}
    (a) Pulse evolution seeded by an external MLL pulse.
    (b) Pulse evolution after initiation using an external cavity and Q-switch.
    (c) Transition from a double-pulse state to a single-pulse state, occurring approximately 340 round-trips after (b).
    (d) Stable pulse train recorded approximately 3000 round-trips after (b).}
    \label{figS:stdynamics}
\end{figure*}
In the experiments, we either use an external mode-locked laser to seed the laser or generate an energetic pulse with an extended cavity and Q-switch to initialize mode-locking.
Meanwhile, several reports have demonstrated self-starting fiber-based Mamyshev oscillators without external seeding~\cite{4591476, Liu24selfstarting, wang2023automated}.
We believe the relative difficulty in starting the integrated Mamyshev oscillator arises from the large difference between the power of parasitic continuous-wave lasing and the peak power of the mode-locked pulses.
To initiate mode-locking, a high intracavity gain is required to amplify small intensity fluctuations into a viable pulse seed.
However, with a typical estimated back-reflection level $10\log_{10}R_\mathrm{parasitic} \approx -27$ dB, parasitic lasing between one of the waveguide Bragg gratings (WBGs) and the reflection point clamps the single-pass gain to only 13.5\,dB, insufficient to amplify noise to level capable of sustained pulsing.
In fiber systems, splicing or using angled end-face connectors can reduce return loss to higher than 50\,dB, making self-starting significantly easier.
Nevertheless, using angled edge couplers and matched fiber array units can significantly reduce chip facet reflectivity, currently the dominant source of parasitic back-reflection.
This reduction may allow for simpler mode-locking initiation in future implementations.

To gain insight into the pulse build-up dynamics, we recorded the laser output using a photodetector and oscilloscope during pulse initiation, either by seeding with an external MLL or by using the extended cavity Q-switching method.
Two distinct behaviors were observed.
Figure~\ref{figS:stdynamics}(a) shows the temporal evolution when the laser is seeded by a short pulse from an external MLL, as described in Section~\ref{ss:seeding}.
In this case, similar to the ``coherence memory'' regime reported in~\cite{Cao24}, the injected seed pulse directly evolve into the stable mode-locked pulse in the cavity.
The pulse amplitude exhibits only a modest $\sim$20\% overshoot before settling into a steady state within approximately 20 round-trips.

Alternatively, pulse initiation can be achieved using an extended cavity arm and electro-optic Q-switch, similar to the scheme in \cite{Sidorenko18} with the effective saturable absorber replaced with a modulator.
In this case, the dynamics differ significantly: mode-locked pulse trains emerge following a high-energy Q-switched pulse, as shown in Figure~\ref{figS:stdynamics}(b).
The Q-switched pulse lasts approximately 24~ns and carries an on-chip energy on the order of 100~nJ, estimated roughly by integrating the pulse area (with possible large error due to photodiode saturation), comparable to the values reported in~\cite{singh2024silicon}.
A train of fast pulses builds up after the slow Q-switched pulse and rapidly increases in amplitude, eventually splitting into two pulses after roughly 150~ns due to excess energy.
This later evolves into a single-pulse state (Fig.\,\ref{figS:stdynamics}(c)) when the population inversion decreased.
Significant amplitude fluctuations are observed immediately after initiation, but they damp out quickly and nearly vanish after around 3000 round-trips (Fig.\,\ref{figS:stdynamics}(d)), leading to a low noise mode-locked state.
This behavior may correspond to the high-energy chaotic state during pulse build up as discussed in~\cite{Cao24}.
\FloatBarrier
\section{Fiber to chip coupling and edge coupler design}
\label{ss:coupling}
\begin{figure*}[htbp]
    \centering
    \includegraphics[width=0.7\linewidth]{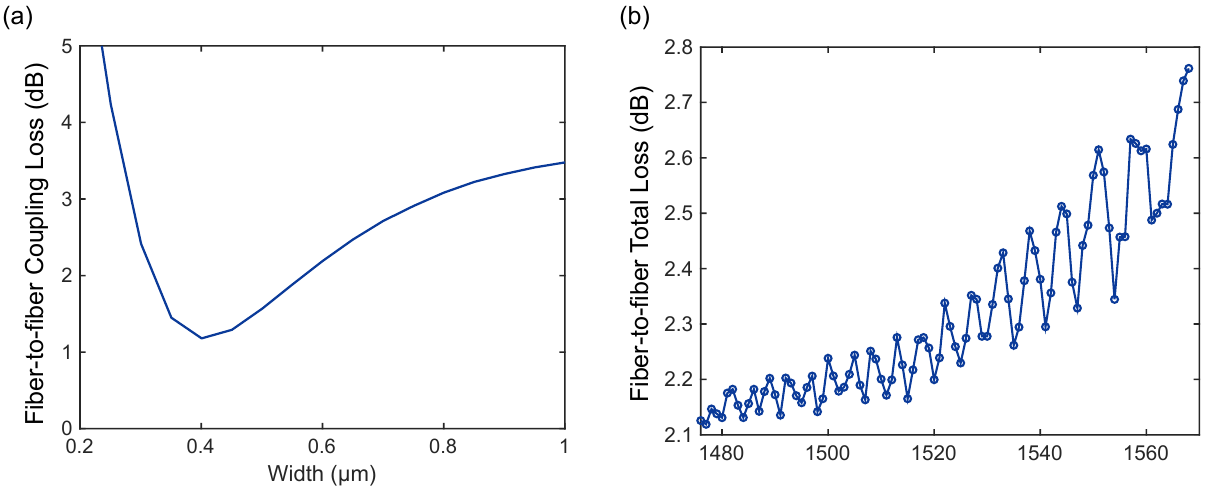}
    \caption{\textbf{Fiber to waveguide coupling loss.} (a) Simulated fiber to chip to fiber coupling loss of TE mode of a cladded 350~nm thick \ce{Si3N4} waveguide to the UHNA7 fiber for different taper widths at \SI{1.55}{\micro\meter}.
    (b) Measured loss from fiber to chip to fiber for a 2~cm reference bus waveguide. }
    \label{figS:coupling}
\end{figure*}

To couple light efficiently from fiber to photonic integrated circuits, we used edge couplers with inverse tapers designed to match the mode field with the fiber used.
Here we use the Coherent UHNA7 fiber that has a mode field diameter (MFD) around \SI{3}{\micro\meter} near 1550~nm wavelength, spliced with a conventional SMF-28 fiber.
The edge coupling efficiency $\eta$ is calculated by the vectorial field overlap \cite{wumodeselective2016}:
\begin{equation}
\eta = \frac{\int \boldsymbol{E}_{\mathrm{f}} \times \boldsymbol{H}_{\mathrm{w}} \cdot \mathrm{d}\boldsymbol{S} \, \int \boldsymbol{E}_{\mathrm{w}} \times \boldsymbol{H}_{\mathrm{f}} \cdot \mathrm{d}\boldsymbol{S}}{\int \boldsymbol{E}_{\mathrm{f}} \times \boldsymbol{H}_{\mathrm{f}} \cdot \mathrm{d}\boldsymbol{S} \, \int \boldsymbol{E}_{\mathrm{w}} \times \boldsymbol{H}_{\mathrm{w}} \cdot \mathrm{d}\boldsymbol{S}},
\end{equation}
where $\boldsymbol{E}_{\mathrm{f}}, \boldsymbol{H}_{\mathrm{f}}, \boldsymbol{E}_{\mathrm{w}}, \boldsymbol{H}_{\mathrm{w}}$ are the electric and magnetic fields of the fiber and waveguide, respectively.
The mode fields are simulated with commercial COMSOL software using the finite element method. 
The parameters of the UHNA7 fiber are extracted from the datasheet.
The waveguide height is designed to be 350~nm as mentioned to leverage a high effective Kerr nonlinearity $\gamma$, while still maintains sufficient overlap between the erbium ion distribution and the optical mode field.
We conservatively designed the length of the tapering section to be \SI{500}{\micro\meter} long to ensure adiabatic mode conversion.
The theoretical coupling efficiency $\eta$ as a function of different taper width is shown in Fig.\,\ref{figS:coupling}(a). The taper width is designed to be around 350~nm to maximize the mode-field overlap, which gives 1.45~dB fiber-to-fiber loss by the simulation.

We measured the coupling loss with a 2~cm dummy straight waveguide from the same wafer without ion implantation by scanning a single frequency laser.
As shown in Fig.\,\ref{figS:coupling}(b), the average total fiber-to-fiber loss between 1540~nm and 1570~nm is approximately 2.52~dB, corresponding to a per-facet coupling loss of 1.26~dB.
This value includes contributions from waveguide propagation loss ($<0.2$~dB) and splice loss between SMF and UHNA7 fibers at both ends.
The splice loss, estimated at 0.62~dB, was characterized by measuring the total loss of three spliced segments (SMF–UHNA7–SMF).

We note that thermal drift and the resulting hysteresis in the coupling setup (despite the use of a TEC) make manual coupling significantly more difficult when the fiber is ``hot'' (carrying significant pump and output power), compared to coupling to dummy waveguides at low power.
The coupling loss described above should therefore be considered as a lower bound for the actual loss during the experiment.

\section{Time domain pulse reconstruction from dispersion sweep data}
It is well known that one cannot retrieve the optical pulse profile unambiguously from the intensity autocorrelation measurement alone. 
In our experiments, since we have acquired a series of intensity autocorrelation $\text{IAC}(t)$ measured under different group delay dispersion values $\beta_2$, together with the measured spectrum intensity $|E(\omega)|^2$, we can computationally reconstruct the amplitude and phase of the pulse.
The phase to retrieve $\phi(\omega)$ is defined by $E(\omega)=|E(\omega)|e^{i\phi(\omega)}$ and its relation to the intensity autocorrelation $\text{IAC}(t)$ can be established by the following equations.
First, the time-domain intensity of the pulse $I(t)$ with quadratic spectral phase assignment ($\beta_2$) is given by:
\begin{equation}
    I(t)=|\mathcal{F}^{-1}\{|E(\omega)|e^{i\phi(\omega)}e^{i\frac{\beta_2}{2}\omega^2}\}|^2,
    \label{Eq:PR_1}
\end{equation}
where $\mathcal{F}^{-1}$ denotes the inverse Fourier transform, and $\omega$ is the offset angular frequency with respect to the carrier frequency of light. 
The autocorrelation of intensity can be computed in the spectral domain using the Wiener–Khinchin theorem:
\begin{equation}
    \text{IAC}(t)=\mathcal{F}^{-1}\{{|\mathcal{F}\{I(t)\}|^2}\},
    \label{Eq:PR_2}
\end{equation}
where $\mathcal{F}$ denotes the Fourier transform. Equations \eqref{Eq:PR_1} and \eqref{Eq:PR_2} describe the forward model of the phase retrieval problem. 

\begin{figure*}[htbp]
    \centering
    \includegraphics[width=1\linewidth]{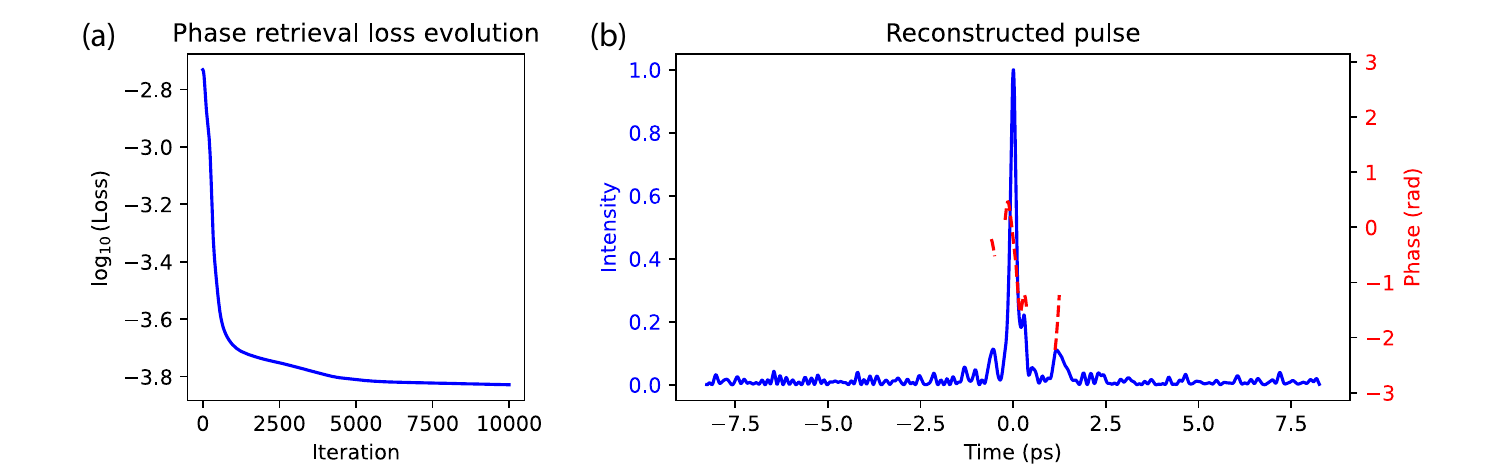}
    \caption{(a) Evolution of the retrieval error over the course of the optimization iterations.  
    (b) Reconstructed pulse intensity and phase at the point of minimum pulse duration, corresponding to a dispersion value of $\beta_2 = 0.0189~\mathrm{ps^2}$.  
    The phase is shown only in regions where the pulse intensity exceeds a threshold for clearness.}
    \label{figS:loss_retrieved_pulse}
\end{figure*}

\begin{figure*}[htbp]
    \centering
    \includegraphics[width=0.7\linewidth]{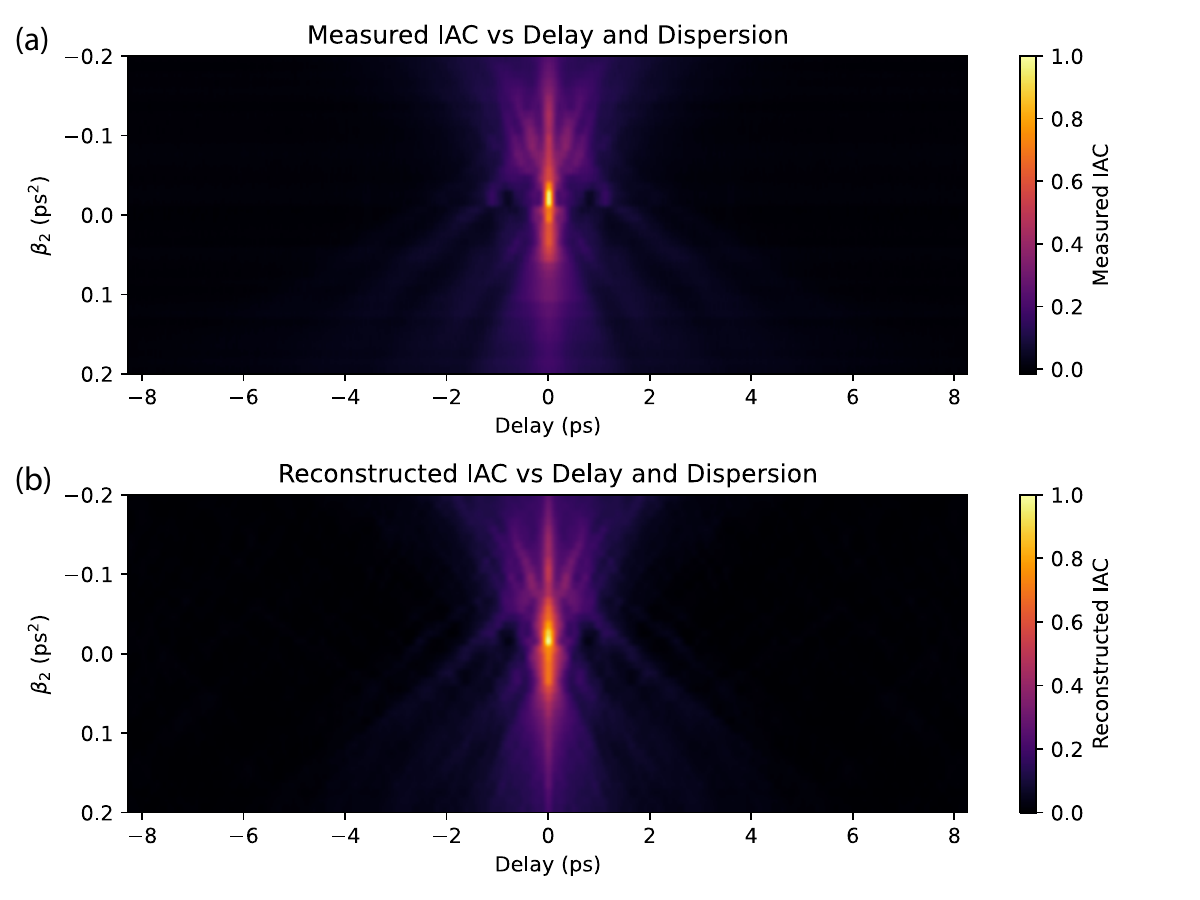}
    \caption{(a) Measured and (b) reconstructed intensity autocorrelation function, shown as a function of dispersion $\beta_2$ and delay.}
    \label{figS:iac_comparison}
\end{figure*}

We use the stochastic gradient descent optimizer in Pytorch, which implements automatic differentiation to calculate the gradient and perform the backpropagation. 
The detailed algorithm is shown in Algorithm \ref{alg:phase_retrieval}.
Notably, we adopted the cosine similarity as the metric for the similarity between the acquired correlation function and the computed correlation function, which is scale-invariant and thus robust to the amplitude mismatches between the computed and measured intensity autocorrelation function.
We also applied a weighted time window to emphasize the region near the zero delay.
In addition, we observed that subtracting the background at large time delays in both measured intensity autocorrelation and the forward model improves the retrieval quality. 
During optimization, the loss converges after 10000 iterations, and the reconstructed pulse reaches the shortest pulse width at $\beta_2=0.0189~{\mathrm{ps^2}}$ (Fig. \ref{figS:loss_retrieved_pulse}). 
The linear temporal phase in the main pulse peak indicates the majority of the pulse energy is well compressed.
Minor ripples in the reconstructed pulse intensity away from the main peaks may be artifacts from the retrieval process.
Figure \ref{figS:iac_comparison} confirms the validity of the retrieved pulse, where the measured intensity autocorrelation is in good agreement with the computed intensity autocorrelation based on the retrieved pulse.

\begin{algorithm}[H]
\caption{Spectral phase retrieval from measured intensity autocorrelation}\label{alg:phase_retrieval}

\KwIn{Measured optical spectrum $|E(\omega)|^2$, measured autocorrelation map $\text{IAC}_{\text{meas}}(t,\beta_2)$ as a function of time delay $t$ and dispersion values $\beta_2$}

\textbf{A. Data preprocessing}\\
Define uniform time and frequency grids\;
Interpolate $\text{IAC}_{\text{meas}}(t,\beta_2)$ and $|E(\omega)|$ onto the uniform grids\;
Normalize, subtract the background, and apply weighting of $\text{IAC}_{\text{meas}}(t,\beta_2)$ for each $\beta_2$: \\
\Indp $\text{IAC}_{\text{meas'}}(t,\beta_2)\leftarrow w(t)\,\bigl(\text{IAC}_{\text{meas}}(t,\beta_2)-\min_t(\text{IAC}_{\text{meas}}(t,\beta_2))\bigr)$\;
\Indm

\textbf{B. Parameter initialization}\\
Set an initial guess for spectral phase $\phi(\omega)$\;
Create a weighting function $w(t)$ centered at zero delay\;

\textbf{C. Forward model}\\
\SetKwProg{Fn}{Function}{}{}\SetKwFunction{FwdModel}{ForwardModel}
\Fn{\FwdModel{$\phi(\omega),\;\beta_2$}}{
    \tcp{Compute the autocorrelation function for a given spectral phase and dispersion.}
    $E(\omega)\leftarrow |E(\omega)|\,e^{i\phi(\omega)}$\;
    $E(t,\beta_2)\leftarrow \mathcal{F}^{-1}\{E(\omega)e^{i\frac{\beta_2}{2}\omega^2}\}$\;
    $I(t,\beta_2)\leftarrow |E(t,\beta_2)|^2$\;
    Compute $\text{IAC}_{\text{fwd}}(t,\beta_2)$ of $I(t,\beta_2)$ via the Wiener–Khinchin theorem\;
}

\textbf{D. Spectral phase retrieval}\\
\For{$k \leftarrow 1$ \KwTo $n_{\text{iter}}$}{
    \ForEach{dispersion value $\beta_2$}{
        Compute the autocorrelation function: $\text{IAC}_{\text{fwd}}(t,\beta_2)\leftarrow \text{\textsc{ForwardModel}}(\phi(\omega),\;\beta_2)$\;
        Subtract background and apply weighting:
        $\text{IAC}_{\text{est}}(t,\beta_2)\leftarrow w(t)\,\bigl(\text{IAC}_{\text{fwd}}(t,\beta_2)-\min_t(\text{IAC}_{\text{fwd}}(t,\beta_2))\bigr)$\;
        Compute cosine similarity loss:
        $L(\beta_2) \leftarrow 1 - \text{cos\_sim}\bigl(\text{IAC}_{\text{est}}(t,\beta_2),\,\text{IAC}_{\text{meas'}}(t,\beta_2)\bigr)$\;
    }
    Compute average loss over all $\beta_2$:
    $L \leftarrow \text{mean}\bigl(L(\beta_2)\bigr)$\;
    Update $\phi(\omega)$ with stochastic gradient descent algorithm\;
}
\end{algorithm}

\FloatBarrier
\section{Simulation and additional experimental results of supercontinuum generation}
\label{ss:supercontinuum}
In this section, we present simulations of supercontinuum generation based on the nonlinear Schr\"odinger equation (NLSE) \cite{dudley2006supercontinuum,guo2018mid}. 
The input pulse is taken from the simulated output of the Mamyshev oscillator after chirp compensation (Fig.\,\ref{figS:tdcompress}(a)).
While the simulated pulse in Fig.\,\ref{figS:tdcompress}(a) is compressed to the minimal width, in practice, the connecting optical fiber may slightly under- or over-compensate the chirp, since pulse compression was experimentally optimized by adjusting the fiber length in $\sim$1~m steps.
To account for this, we also include a small residual chirp in the simulated input pulse.
The on-chip average power was set to approximately 18~mW (corresponding to a peak power of $\sim$450~W), based on experimental estimates that take into account the input coupling loss.
The nonlinear coefficient was calculated as a function of wavelength using the simulated effective mode area, and a propagation loss of 5~dB/m was assumed for the waveguide.
The dispersion of the \ce{Si3N4} waveguide (Fig.~4(e) in the main text) was extracted from the broadband mode effective index simulation, obtained in COMSOL Multiphysics with the finite element method. 
The simulated waveguide cross-section was set to \SI{2.07}{\micro\metre} $\times$ \SI{0.70}{\micro\metre}, based on SEM measurements (Fig.~4(c) in the main text).

\begin{figure*}[htbp]
    \centering
    \includegraphics[width=0.75\linewidth]{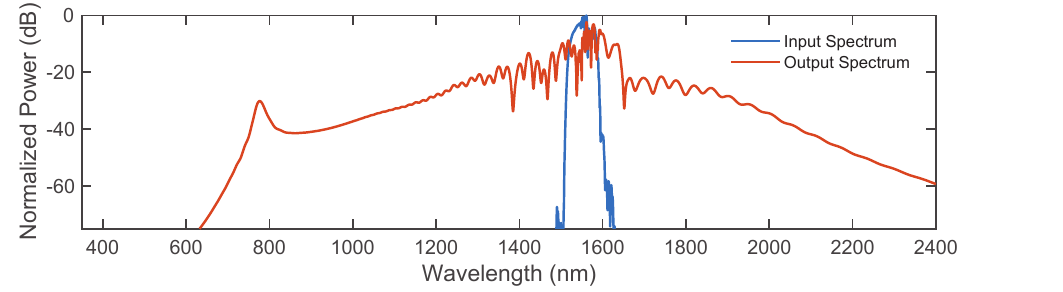}
    \caption{Simulated supercontinuum generated in a 43.7~mm \ce{Si3N4} waveguide driven by the Mamyshev oscillator.}
    \label{figS:simulated_sc_short}
\end{figure*}

\begin{figure*}[htbp]
    \centering
    \includegraphics[width=0.75\linewidth]{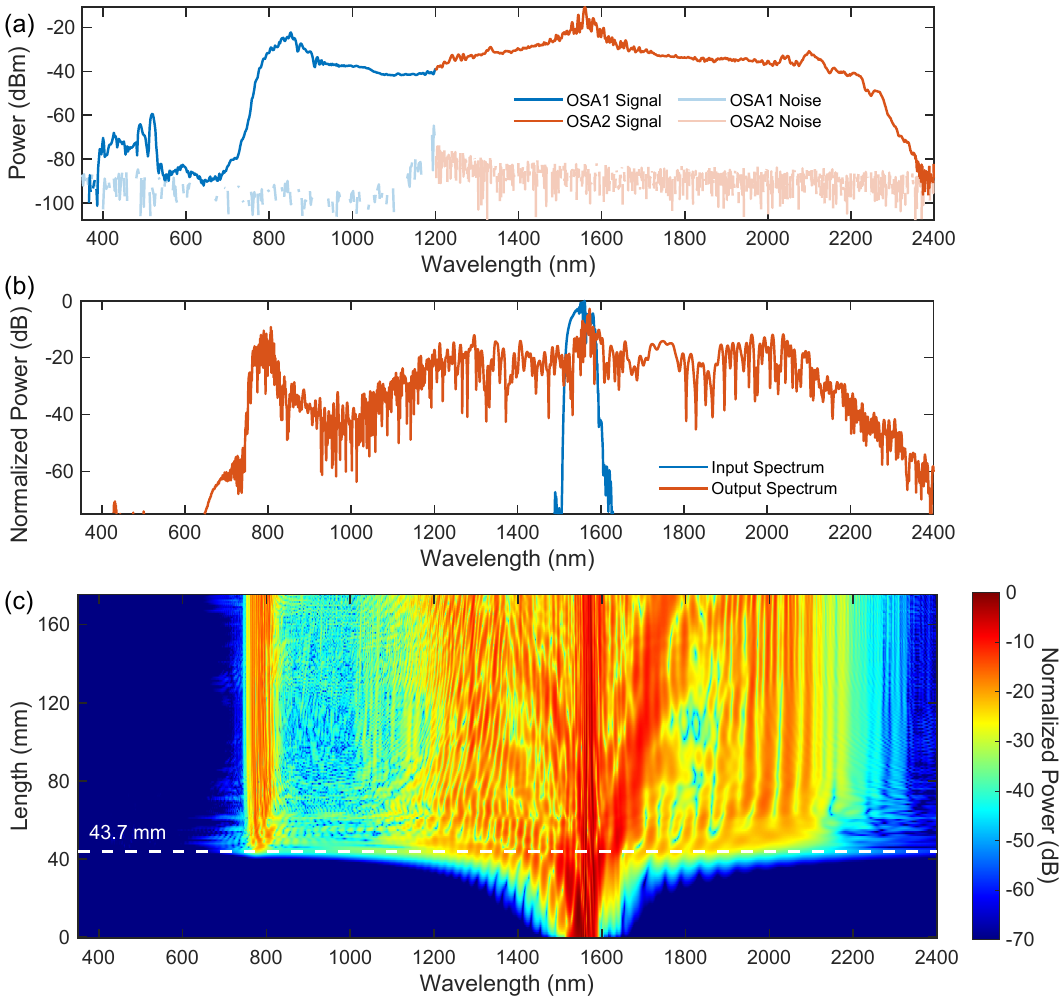}
    \caption{(a) Experimentally measured supercontinuum generation in a 175~mm \ce{Si3N4} waveguide.
    (b) Simulated supercontinuum generation in a 175~mm \ce{Si3N4} waveguide driven by the Mamyshev oscillator.
    (c) Simulated evolution of spectra over a propagation distance of 175~mm.
    The dashed line marks the propagation distance of 43.7~mm, corresponding to the spectrum in Fig.\,\ref{figS:simulated_sc_short}.}
    \label{figS:simulated_sc_long}
\end{figure*}

Figure~\ref{figS:simulated_sc_short} shows the simulated supercontinuum spectrum in a 43.7~mm dispersion-engineered \ce{Si3N4} waveguide, exhibiting good agreement with the measured spectrum (Fig.~4(d) in the main text).
Minor discrepancies may arise from uncertainties in the \ce{Si3N4} refractive index or variations in waveguide height across different samples.
We also experimentally investigated supercontinuum generation in a longer \ce{Si3N4} waveguide (175~mm) with the same cross-section.
The corresponding measured and simulated spectra are presented in Fig.\,\ref{figS:simulated_sc_long}(a) and (b), respectively.
Compared to the 43.7~mm device, the 175~mm waveguide shows enhanced spectral broadening toward both shorter and longer wavelengths, though likely at the expense of reduced coherence.
Figure~\ref{figS:simulated_sc_long}(c) illustrates the simulated spectral evolution along the propagation length.
The results indicate that the 43.7~mm waveguide length approximately corresponds to the onset of octave-spanning supercontinuum generation, with the initial section contributing to compensating for the residual chirp of the input pulse.

\section{Discussion on the fabrication process}
\label{ss:fab}
The current fabrication process exhibits two known issues, both of which can be readily addressed.
First, the highly-stressed \SI{350}{\nano\metre} \ce{Si3N4} layer is prone to cracking within days if not promptly patterned, significantly reducing device yield.
While these cracks can be small and difficult to detect even under an optical microscope before annealing, they are enough to destroy the waveguides, especially after annealing-induced shrinkage of the \ce{Si3N4}.
In later fabrication runs, we demonstrated that patterning deep trenches (\SI{2}{\micro\metre} wide, square grid, $\sim$\SI{3}{\micro\metre} deep) into the oxide layer along the wafer edge, similar to those described in reference~\cite{ji2024efficient}, effectively prevents cracking.

Second, the high-dose ion implantation causes swelling at the top of the \ce{Si3N4} waveguides, creating an overhanging structure at the edges.
These edges can lead to formation of voids during ICPCVD deposition of the \ce{SiO2} cladding, despite enhanced corner coverage from the optimized process.
Such voids often have irregular geometry and high refractive index contrast with \ce{SiO2} cladding, increasing waveguide scattering loss.
This issue can be mitigated by introducing a lateral cladding prior to ion implantation.
For example, wafers before implementation can be fabricated via the photonic damascene process~\cite{liu2021high,liu2022photonic} or through LPCVD cladding deposition followed by chemical-mechanical polishing and controlled \ce{HF} thinning.

\bibliographystyle{naturemag}
\bibliography{ref}